\documentclass[11pt]{article}
\usepackage{bm,bbm}
\usepackage{amsmath,amssymb,indentfirst}
\usepackage{graphicx}           
\usepackage{time}
\usepackage{indentfirst}                
\usepackage{makeidx}                    
\usepackage{listings}                   
\usepackage{booktabs}
\usepackage[round,sort,nonamebreak]{natbib} 
\usepackage{natbib}                      
\usepackage{multirow}
\usepackage{lineno}
\usepackage[pdftex,pagebackref=true]{hyperref} 
\usepackage[usenames,dvipsnames]{color} 
\usepackage{listings}
\graphicspath{{./figuras/}}             

\date{}

\begin{document}


\title{Diagnostic tools for a multivariate negative binomial model for fitting correlated data with overdispersion}
\author
{
Lizandra Castilho Fabio \\
Department of Statistics\\
Federal University of Bahia\\
Brazil
\and
Cristian Villegas\\
Department of Exact Sciences\\
University of São Paulo\\
Brazil
\and
Jalmar M. F. Carrasco{\footnote{Corresponding author. Email: carrascojalmar@gmail.com}}\\
Department of Statistics\\
Federal University of Bahia\\
Brazil
\and
M\'ario de Castro\\
Instituto de Ciências Matemáticas e de Computação \\
Universidade de São Paulo\\
Brazil
}

\maketitle
\begin{abstract}
We focus on the development of diagnostic tools and an \texttt{R} package called \texttt{MNB} for a multivariate negative binomial (MNB) regression model for detecting atypical and influential subjects. The MNB model is deduced from a Poisson mixed model in which the random intercept follows the generalized log-gamma (GLG) distribution. The MNB model for correlated count data leads to an MNB regression model that inherits the features of a hierarchical model to accommodate the intraclass correlation and the occurrence of overdispersion simultaneously. The asymptotic consistency of the dispersion parameter estimator depends on the asymmetry of the GLG distribution. Inferential procedures for the MNB regression model are simple, although it can provide inconsistent estimates of the asymptotic variance when the correlation structure is misspecified. We propose the randomized quantile residual for checking the adequacy of the multivariate model, and derive global and local influence measures from the multivariate model to assess influential subjects. Finally, two applications are presented in the data analysis section. The code for installing the \texttt{MNB} package and the code used in the two examples is exhibited in the Appendix.

\vspace{0.5cm}

\noindent \textbf{keywords:} Count data; Overdispersion; Multivariate negative binomial distribution; \texttt{MNB} package.

\end{abstract}

\section{Introduction}

Hierarchical models have been suggested for analyzing correlated count data due to their feature of allowing one to model the intraclass correlation and accommodate the occurrence of overdispersion simultaneously. This is handled through the inclusion of random effects in the systematic component of generalized linear models. Using this approach, it is possible to relax the assumption about the distribution of the random effects, taking into account the empirical distributions of the data or individual profiles (\citealp{Pawitan06}, \citealp{Molenberghs07} and \citealp{Fabio12}).
Several methods have been proposed for inferential procedures in hierarchical models due to the intractable integrals involved in inference functions. The latter fact has been an obstacle in the development of diagnostic tools, leading to completely numerical procedures being used. 
In such cases, a multivariate model deduced from the random effects approach (\citealp{Mol05}) can be an alternative to a hierarchical model for fitting correlated count data with extra variability. \cite{Fabio12} proposed the random intercept Poisson mixed regression model by assuming that the random effects follow a generalized log-gamma (GLG) distribution (\citealp{Lawless87}). 
This distribution can be skewed to the right or skewed to the left, with the normal distribution as a particular case. Thus, the random intercept Poisson-GLG model reduces to a multivariate negative binomial (MNB) model when it is assumed that the scale and shape parameters of the GLG distribution are equal. The random intercept of a Poisson-GLG model is able to accommodate intraclass correlation and handle overdispersion, due to the several degrees of asymmetry that the GLG distribution can assume. The MNB inherits these features, with the overdispersion parameter $\phi^{-1} = \lambda^{2}$ and simpler correlation structures. In the literature, the MNB regression (MNBR) model
 has been used for modeling correlated count data from several areas, for instance,  health  
(\citealp{Solis05}), spatial (\citealp{Moller2010}) and economics (\citealp{Sung2018}) data. 

We focus on developing diagnostic tools and an \texttt{R} \citep{R2017} package for the MNBR model under overdispersion, which are essential for detecting outliers and for checking the adequacy of the model 
(\citealp{cook+weisberg}). Simulation studies are also conducted to evaluate the asymptotic properties of the maximum likelihood (ML) estimator and to analyze how these properties’ are impacted on the random effect distribution misspecification. As noted in \citet{Solis05}, the MNBR model can provide inconsistent estimates of the asymptotic variance of the regression coefficient when the covariance matrix of the random effect is misspecified. 
We use 
randomized quantile residuals (\citealp{Dunn+Smyth}) for checking the adequacy of the MNBR  model. 
Following \citet{Cook77,Cook} approach, global and local influence measures are  developed from the MNBR model to detect influential subjects. These methodologies are useful to assess the impact on the estimation procedure by removing the subject from the data set and by proposing perturbation models, respectively. The total local influence measure suggested by \citet{Lesafre} is also extended for the MNBR model. These measures are helpful for identifying outlying subjects that matter in the MNBR model and interpret them according to the application. 
The inferences and diagnostic analysis from MNBR are performed by using the \texttt{MNB} package developed by the authors.  

The paper is organized as follows: In Section \ref{MNBD}, we provide some background about the MNBR model approach by \cite{Fabio12}. In Section \ref{sec:diag}, we discuss diagnostic analysis methodologies for the MNBR model. In Section \ref{sec:simu}, we perform a simulation study to evaluate the asymptotic behavior of the ML estimator.
In Section \ref{apli}, the diagnostic analysis is applied to two real data sets, using the \texttt{MNB} package. The code for installing the \texttt{MNB} package is presented in the Appendix. Finally, in Section \ref{conclu}, we provide some discussions. 
 
\section{MNBR model}
\label{MNBD}

Let $y_{ij}$ denote the $j$th measurement taken on the $i$th subject or cluster, for $j=1,\ldots,m_i$ and $i=1,\ldots,n$. Further, let $b_i$ be  random effects that follow the GLG distribution \citep{Law03}. 
Assuming that $y_{ij}|b_i$ are independent outcomes with probability mass function represented by a Poisson distribution, \cite{Fabio12} 
proposed a random intercept Poisson-GLG model with 
the following 
hierarchical structure: 
$(i)$~$y_{ij} |b_i$ $\stackrel{\rm ind} {\sim}$ ${\rm Poisson}(u_{ij})$,
$(ii)$~$u_{ij} = \mu_{ij}\exp(b_i),$ and
$(iii)$~$b_{i} \stackrel{\rm iid} {\sim} {\rm GLG}(0, \sigma, \lambda),$ 
where $\mu_{ij}=\exp{(\bm{x}^{\top}_{ij}\bm{\beta}})$,  with  
$\bm{x}_{ij}=(\bm{x}_{ij1},\ldots,\bm{x}_{ijp})^{\top}$ 
containing the values of explanatory variables, $\bm{\beta}= (\beta_{1},\ldots,\beta_{p})^{\top}$ the vector of regression coefficients, and $\sigma>0$ and $\lambda \in \mathbb{R}$ the scale and shape parameters of the GLG distribution. In general, 
in the random effects approach, the marginal distribution of  
the $i$th subject does 
not have an explicit form (\citealp{Mol05} pp. 259-266).
The ML estimates are obtained by integrating out the random effect and maximizing the log-likelihood function.
\cite{Fabio12} showed that the integral can be solved analytically when the scale and shape parameters of the GLG distribution are equal ($\sigma=\lambda$, with $\lambda \in \mathbb{R}^{+}$). 
When $\phi=\lambda^{-2}$, the MNB distribution is deduced from the random intercept Poisson-GLG model of the following form:
\begin{equation}
\label{eq2a}
f(\bm{y}_i;\bm{\beta},\phi)=\frac{\Gamma(\phi + y_{i+})\phi^{\phi}}{\left(\prod_{j=1}^{m_i}y_{ij}!\right)\Gamma(\phi)}\frac{\exp\left(\sum_{j=1}^{m_i}y_{ij}\log(\mu_{ij})\right)}{(\phi + \mu_{i+})^{\phi + y_{i+}}},
\end{equation}
where $\bm{y}_i=(y_{i1},\ldots,y_{im_i})^{\top}$ is the vector of measurements available for 
the $i$th subject, $\phi$ is the dispersion parameter, $\Gamma(\cdot)$ is the gamma function,  $y_{i+}=\sum_{j=1}^{m_i} y_{ij}$, and $\mu_{i+}=\sum_{j=1}^{m_i} \mu_{ij}$. 
 According to \cite{Kotz97}, the MNB distribution belongs to the
discrete multivariate exponential family of distributions, and 
its marginals $f(y_{ij})$ are negative binomial distributions
 with $\textrm{E}(y_{ij}) = \mu_{ij}$ 
and $\textrm{Var}(y_{ij})= \mu_{ij} + \mu_{ij}^2 / \phi$, for $i=1,\ldots,n$ and $j=1,\ldots,m_i.$ 
 The covariances $\textrm{Cov}(y_{ij},y_{ij^{'}}) = \mu_{ij} \mu_{ij^{'}} / \phi$ and intraclass correlations 
$\textrm{Corr}(y_{ij},y_{ij^{'}})=\sqrt{\mu_{ij}\mu_{ij^{'}}}/\left(\sqrt{\phi + \mu_{ij}}\sqrt{\phi + \mu_{ij^{'}}}\right)$ for $j \neq j^{'}$ are always positive.
When $\phi^{-1}$ indicates the amount of excess correlation in the data (\citealp{Hilbe11}), the  parameter $\phi$ is also called an overdispersion parameter.
For large values of $\phi$, the marginals of the MNB distribution behave approximately as independent Poisson distributions with mean $\mu_{ij}$. By $\bm{y}_i \sim \textrm{MNB}(\bm{\mu}_i, \phi),$ we denote independent vectors of random outcomes
 that follow the probability function given in (\ref{eq2a}), with $\bm{\mu}_i=(\mu_{i1},\ldots, \mu_{im_i})^{\top}$ and $\phi>0$.

The MNBR model is defined by assuming $(i)$~$\bm{y}_{i}\stackrel{\rm ind} {\sim}{\rm MNB}(\bm{\mu_{i}},\phi)$ and  
$(ii)$~$\log(\mu_{ij}) = \bm{x}_{ij}^{\top}\bm{\beta}$. 
Letting $\bm{y}=(\bm{y}^{\top}_{1},\ldots,\bm{y}^{\top}_{n})^{\top}$ be the vector containing  all the measured outcomes for the $i$th subject, the log-likelihood function is given by
\begin{align}
\label{lvero}
\ell(\bm{\theta}) & =  \sum_{i=1}^{n}\log\left\{\frac{\Gamma(\phi + y_{i+})}{\Gamma(\phi)}\right\}   -
                    \sum_{i=1}^{n}\sum_{j=1}^{m_i}\log \left(y_{ij}!\right) + n\phi\log(\phi) \nonumber \\
& \quad - \phi\sum_{i=1}^{n}\log(\phi + \mu_{i+}) +
  \sum_{i=1}^{n}\sum_{j=1}^{m_i}y_{ij}\log\left\{\frac{\mu_{ij}}{\phi + \mu_{i+}} \right\},
\end{align}
where $\bm \theta=(\bm \beta^{\top},\phi)^{\top}$. The ML estimates 
$\widehat{\bm \theta}$ of $\bm \theta$ are computed by using the quasi-Newton (\texttt{BFGS}) method. Alternatively, we can solve the nonlinear equation obtained by setting the components of the score vector equal to zero, that is, $\bm{U}_{\theta}=(\bm{U}_{\beta}^{\top}, U_{\phi})^{\top}=\bm{0}$ (details about the calculations are in the Appendix).  For interval estimation and hypothesis tests on the model parameters, the expected or observed Fisher information is required. 
Under standard regularity conditions, 
$\widehat{\bm{\theta}} - \bm{\theta}$ is asymptotically distributed as a multivariate normal distribution with mean $\bm{0}$ and covariance matrix equal to the inverse of the Fisher information matrix of the MNBR model (see \citealp{Fabio12}).
	
\section{Diagnostic analysis}
\label{sec:diag}

This section is devoted to diagnostic tools for the MNBR model. In what follows, we will discuss the residual analysis, and the global and local influence methodologies.

\subsection{Residual analysis}
\label{subsec:residuos}

Let $\bm{y}_i=(y_{i1}, y_{i2}, \ldots, y_{im_{i}})^{\top}$ be a random vector that follows a MNB distribution. According to \cite{Tsui}, the distribution of $y_{i+}=\sum_{j=1}^{m_i}y_{ij}$ is negative binomial with probability distribution function given by
\begin{eqnarray*}
f(y_{i+})=\frac{\Gamma(y_{i+}+\phi)}{(y_{i+})! \Gamma(\phi)}q^{\phi}(1-q)^{y_{i+}}, \ \ \ y_{i+}=0,1,2, ...,
\end{eqnarray*}
where $q=\phi/(\phi+\mu_{i+})$. Then, the randomized quantile residuals (\citealp{Dunn+Smyth}), which follow a standard normal distribution, can be used to assess departures from the MNBR model. If $F(y_{i+}; \mu,\phi)$ is the cumulative distribution of $f(y_{i+})$, $a_{i}=\lim_{y \uparrow y_{i+}} F(y; \widehat{\mu_i},\widehat{\phi})$, and $b_{i}=F(y_{i+}; \widehat{\mu}_i,\widehat{\phi})$, then the randomized quantile residuals for $\bm{y}_i$ are given by $r_{q,i}=\Phi^{-1}(u_i)$,
where $\Phi(\cdot)$ is the cumulative distribution function of the standard normal and $u_{i}$ is a uniform random variable on the interval $(a_i,b_i]$. 

\subsection{Global influence}

Based on the case-deletion approach (\citealp{Cook77}), we propose the generalized Cook's distance  as a global influence measure to assess the impact on the ML estimates of the MNBR model when the $i$th subject (or cluster) is removed from the data set. Let $\bm{y}_{(i)}$ be a vector of random outcomes after deleting the $i$th subject. Let $\widehat{\bm{\theta}}_{(i)}$ be the ML estimate of $\bm{\theta}$ computed from $\ell_{(i)}(\bm{\theta})$, where $\ell_{(i)}(\bm{\theta})$ is obtained from (\ref{lvero}) after removing the $i$th subject. The generalized Cook's distance is  
defined as the standardized norm of the distance between $\widehat{\bm{\theta}}_{(i)}$ and $\widehat{\bm{\theta}}$, given by the expression
\begin{equation*}
\label{cook}
GD_{i}(\bm{\theta})=(\widehat{\bm{\theta}}_{(i)}-\widehat{\bm{\theta}})^{\top}
[\ddot{\ell}_{\bm{\theta} \bm{\theta}}]^{-1}
(\widehat{\bm{\theta}}_{(i)}-\widehat{\bm{\theta}}),
\end{equation*}
where $\ddot{\ell}_{\bm{\theta} \bm{\theta}}$ (see the Appendix) is the Fisher information matrix of the MNBR model. Large values of $GD_i$ indicate that the ML estimates are strongly influenced by deleting the $i$th subject. Another popular measure of the difference between $\widehat{\bm{\theta}}_{(i)}$ and $\widehat{\bm{\theta}}$ is the likelihood displacement given by $LD_{(i)}(\bm{\theta})=2\{\ell(\widehat{\bm{\theta}})-\ell(\widehat{\bm{\theta}}_{(i)})\}$.

\subsection{Local influence}

The local influence methodology was proposed by \cite{Cook} for assessing the sensitivity of the parameters estimated when small perturbations 
are introduced in the model.
Consider the perturbation vector $\bm{\omega} = (\omega_1, \ldots,\omega_v)^{\top}$, varying in some open subset $\Omega \subset \mathbb{R}^{v}$. Let  $\ell(\bm{\theta}|\bm{\omega})$ denote the log-likelihood function of the perturbed model. It is assumed there exists $\bm{\omega}_0 \in \Omega$  such that 
$\ell(\bm{\theta}|\bm{\omega}_0)= \ell(\bm{\theta})$ for all $\bm{\theta}.$ 
The influence of the minor perturbation on the ML estimate 
$\widehat{\bm{\theta}}$ 
may be assessed by the likelihood displacement   
$LD(\bm{\omega})=2\{\ell(\widehat{\bm{\theta}})-\ell(\widehat{\bm{\theta}}_{\bm{\omega}})\}$, 
where $\widehat{\bm{\theta}}_{\bm{\omega}}$ denotes the ML estimate from the perturbed model. A plot of $LD(\bm{\omega})$ versus $\bm{\omega}$ contains essential information about the
influence of a perturbation scheme.
Cook's idea consists of selecting a unit direction $\bm{d}$ and evaluating the plot $LD(\omega_{0}+a\bm{d})$, where $a \in \mathbb{R}$. This plot is called lifted line, and each fitted line can be obtained by considering the normal curvature defined by
$C_{\bm{d}}(\bm{\theta})=2|\bm{d}^{\top}\bm{\Delta}^{\top}\ddot{\ell}(\bm{\theta})^{-1} \bm{\Delta}\bm{d}|$ around $a=0$, where $\ddot{\ell}(\bm{\theta})=\partial^{2}\ell(\bm{\theta})/\partial \bm{\theta}\partial{\bm{\theta}}^{\top}$  is evaluated at $\widehat{\bm{\theta}}$ and 
$\bm{\Delta}=\partial^{2}\ell(\bm{\theta}|\bm{\omega})/\partial \bm{\theta}\partial \bm{\omega}^{\top}$   is evaluated at $\widehat{\bm{\theta}}$ and $\bm{\omega}_{0}.$ 
Large values of $C_{\bm{d}}(\bm{\theta})$ indicate the sensitivity  induced by perturbation
schemes in direction $\bm{d}.$ 
\cite{Cook} suggests the local influence measure $C_{\bm{d}_{max}}(\bm{\theta})$, evaluated in the direction $\bm{d}_{max}$ corresponding to the eigenvector with maximal normal curvature. 
If the $i$th component of $\bm{d}_{max}$  is relatively large, this indicates that perturbations may lead to substantial changes in the ML estimates. Based on this approach, \cite{Lesafre} suggested the total local curvature corresponding to the $i$th element.  This local influence measure is obtained by taking the direction $\bm{d}_{i}$, an $n \times 1$ vector of zeros with a 1 in the $i$th position. The total local curvature in direction $\bm{d}_{i}$ assumes the form
$C_i(\bm{\theta})=2|\bm{\Delta}_i^{\top}\ddot{\ell}(\bm{\theta})^{-1} \bm{\Delta}_i|$, where 
$\bm{\Delta}_i=\partial^{2}\ell_i(\bm{\theta}|\bm{\omega})/\partial \bm{\theta}\partial \bm{\omega}^{\top}$ denotes the $i$th row of $\bm{\Delta}$, such that
$\ell_i(\bm{\theta}|\bm{\omega})$ is the contribution of the $i$th individual to the log-likelihood function.
$\bm{\Delta}_i$ is evaluated at 
$\widehat{\bm{\theta}}$ and $\bm{\omega}_0$, and it is recommended to look at the index plot of $C_{i}(\bm{\theta})$ to assess influential subjects. Usually, in the literature, three types of perturbation schemes are considered for count data: case weights, explanatory variable, and dispersion parameter perturbation schemes.

\subsection*{\underline{Case weights perturbation for the $i$th subject}}

The log-likelihood function of the perturbed model takes the form
\begin{align*}
\ell(\bm{\theta}|\bm{\omega})& = \sum_{i=1}^{n}\omega_{i}\Big\{\log\Big( \Gamma(\phi + y_{i+}) \Big) - \log\Big( \Gamma(\phi) \Big) - \sum_{j=1}^{m_i}\log(y_{ij}!) + \phi\log(\phi) \\
                          & \quad - \phi\log(\phi + \mu_{i+}) + \sum_{j=1}^{m_i}y_{ij}\log(\mu_{ij}) - \sum_{j=1}^{m_i}y_{ij}\log(\phi + \mu_{i+})\Big\}.
\end{align*}

The $\bm{\Delta}$ matrix in $C_{\bm{d}}(\bm{\theta})$ is given by
$\bm{\Delta}^{\top}=(\bm{\Delta}^{\top}_{\bm \beta},\bm{\Delta}^{\top}_{\phi})^{\top}$, with
$\bm{\Delta}^{\top}_{\bm{\beta}}=(\bm{\Delta}^{\top}_{\beta_1},\bm{\Delta}^{\top}_{\beta_2},\ldots,\bm{\Delta}^{\top}_{\beta_p})$ and 
$\bm{\Delta}_{\beta_k}=(\Delta_{k 1},\Delta_{k 2},\ldots, \Delta_{k i},\ldots, \Delta_{k n}),$ where 
\begin{align*}
\Delta_{ki} &=\sum_{j=1}^{m_i} \left\{ y_{ij}  -
\frac{(\widehat{\phi} + y_{i+})}{(\widehat{\phi} +
    \widehat{\mu}_{i+})} \exp(\bm{x}^{\top}_{ij}\widehat{\bm{\beta}})
\right\} \bm{x}_{ijk}= \sum_{j=1}^{m_{i}}(y_{ij}-\widehat{a}_{i}\widehat{\mu}_{ij})\bm{x}_{ijk},
\end{align*}
 and $\bm{\Delta}^{\top}_{\phi}=(\Delta_{\phi 1},\Delta_{\phi 2},\ldots,\Delta_{\phi i}, \ldots,\Delta_{\phi n})$, with
\begin{align*}
\Delta_{\phi i} =
\psi(\widehat{\phi} + y_{i+})- \psi(\widehat{\phi})+\log \left( \frac{\widehat{\phi}}{\widehat{\phi} + \widehat{\mu}_{i+}}\right)+
\left(\frac{\widehat{\mu}_{i+}-y_{i+}}{\widehat{\phi} + \widehat{\mu}_{i+}}\right),
\end{align*}
where $\psi(\cdot)$ is the digamma function, $\widehat{\mu}_{i+}=\sum_{j=1}^{m_i}\widehat{\mu}_{ij}$, and $\widehat{\mu}_{ij}=\exp(\bm{x}_{ij}^{\top}\widehat{\bm{\beta}})$, for $j=1,\ldots,m_i$ and $i=1,\ldots,n$. 

\subsection*{\underline{Case weights perturbation for the $j$th measurement of the $i$th subject}}

For this perturbation scheme, the log-likelihood function can be expressed as
\begin{align*}
\ell(\bm{\theta}|\bm{\omega})&=\sum_{i=1}^{n}\sum_{j=1}^{m_{i}}\omega_{ij}\Big\{\frac{1}{m_{i}}\log \Gamma(\phi+y_{i+})-\frac{1}{m_{i}}\log \Gamma(\phi)-\log(y_{ij}!)\\
& \quad{+}\quad\frac{1}{m_{i}}\phi\log(\phi)-\frac{1}{m_{i}}\phi \log(\phi+\mu_{i+})+y_{ij}\log(\mu_{ij})-y_{ij}\log(\phi+\mu_{i+})\Big\}.
\end{align*}

The matrix  
$\bm{\Delta}=(\bm{\Delta}^{\top}_{\bm \beta},\bm{\Delta}_{\phi}^{\top})^{\top},$ in which element $ij$ of $\bm{\Delta}^{\top}_{\bm \beta}$ and $\bm{\Delta}^{\top}_{\phi}$, respectively, is given by
\begin{align*}
\Delta_{k{ij}} &=y_{ij}x_{ijk}-\frac{\widehat{\phi}/m_{i}+y_{ij}}{\widehat{\phi}+\widehat{\mu}_{i+}}\sum_{j=1}^{m_{i}}\widehat{\mu}_{ij}x_{ijk}, k=1,\ldots,p \quad {\rm and}\\
\Delta_{\phi{ij}} &= \frac{1}{m_{i}}(\psi(\widehat{\phi}+y_{i+})-\psi(\widehat{\phi}))+\frac{1}{m_{i}}(1+\log(\widehat{\phi})-\log(\widehat{\phi}+\widehat{\mu}_{i+}))\\
&\quad{-}\quad\frac{\widehat{\phi}/m_{i}+y_{ij}}{\widehat{\phi}+\widehat{\mu}_{i+}}.
\end{align*}

\subsection*{\underline{Explanatory variable perturbation}}

We now consider an additive perturbation on a particular continuous explanatory variable, $\bm{x}_{ijk}, j=1,\ldots,m_{i}, i=1,\ldots,n, k=1,\ldots,p$, by setting $\bm{x}^{*}_{ijk}=\bm{x}_{ijk}+\omega_{i}S_{x}$, where $S_{x}$ is a scale factor and $\omega_{i} \in \mathbb{R}$. This perturbation scheme leads to the following expression for the log-likelihood function: 
\begin{align*}
\ell(\bm{\theta}|\bm{\omega})&=\sum_{i=1}^{n} \Bigg\{ \log\Big( \Gamma(\phi + y_{i+}) \Big) - \log\Big( \Gamma(\phi) \Big) - \sum_{j=1}^{m_i}                            \log(y_{ij}!) + \phi\log(\phi) \\
                          & \quad - \phi\log(\phi + \mu_{i+}^{*}) + \sum_{j=1}^{m_i}y_{ij}\log(\mu_{ij}^{*}) - \sum_{j=1}^{m_i}y_{ij}\log(\phi +                            \mu_{i+}^{*}) \Bigg\},
\end{align*}
where $\mu_{i+}^{*}=\sum_{j=1}^{m_i}\mu_{ij}^{*}$, $\mu_{ij}^{*}=\exp(\bm{x}_{ij}^{*\top}\bm{\beta})$,
$\bm{x}_{ij}^{*\top}\bm{\beta}=\beta_{1}+\beta_{2}x_{ij2}+\ldots+\beta_{k}(x_{ijk}+\omega_{i}S_{x})+\ldots+\beta_{p}x_{ijp}$, and $\bm{\omega}_{0}=(0,\ldots,0)^{\top}$. For $k=1,\ldots,p$, if $t=k$, $\bm{\Delta}_{t}$
is given by
\begin{align*}
\Delta_{ti}&=\frac{\widehat{\beta}_{t}S_{x}\widehat{\phi}\exp(\bm{x}^{\top}_{ij}\widehat{\bm{\beta}})}{(\widehat{\phi}+\widehat{\mu}_{i+})} \Bigg(1-\frac{\widehat{\mu}_{i+}}{\widehat{\phi}+\widehat{\mu}_{i+}}\Bigg)x_{ijt}+\frac{S_{x}\exp(\bm{x}_{ij}^{\top}\widehat{\bm{\beta}})}{\widehat{\phi}+\widehat{\mu}_{i+}}\\
								&
                                                                  \quad + \sum_{j=1}^{m_{i}}y_{ij}\Big[S_{x}+\Big(\widehat{\beta}_{t}S_{x}-\frac{1}{\widehat{\mu}_{ij}}\Big)x_{ijt}\Big]+
                  \sum_{j=1}^{m_{i}}\frac{y_{ij}\exp(\bm{x}_{ij}^{\top}\widehat{\bm{\beta}})}{\widehat{\phi}+\widehat{\mu}_{i+}} \\
								&
                                                                  \quad \times \Big[S_{x}+(\widehat{\beta}_{t}S_{x}-1)x_{ijt}\Big].
\end{align*}
If $t \neq k$,
\begin{align*}
\Delta_{ti}&=\frac{\widehat{\beta}_{t}S_{x}\widehat{\phi}\exp(\bm{x}^{\top}_{ij}\widehat{\bm{\beta}})}{(\widehat{\phi}+\widehat{\mu}_{i+})}\Bigg(1-\frac{\widehat{\mu}_{i+}}{\widehat{\phi}+\widehat{\mu}_{i+}}\Bigg)x_{ijt}+ \sum_{j=1}^{m_{i}}y_{ij}x_{ijt} \\
								&
                                                                  \quad \times \Big(\widehat{\beta}_{t}S_{x}-\frac{1}{\widehat{\mu}_{ij}}\Big)+
                  \sum_{j=1}^{m_{i}}\frac{(\widehat{\beta}_{t}S_{x}-1)}{(\widehat{\phi}+\widehat{\mu}_{i+})}\exp(\bm{x}_{ij}^{\top}\widehat{\bm{\beta}})y_{ij}x_{ijt},
\end{align*}
and the $i$th element of $\bm{\Delta}^{\top}_{\phi}$ is
\begin{equation*}
\Delta_{\phi_i}=\frac{\widehat{\beta}_{t}S_{x}\widehat{\mu}_{i+}}{(\widehat{\phi}+\widehat{\mu}_{i+})}\Big(\frac{\widehat{\phi}}{\widehat{               \phi}+\widehat{\mu}_{i+}}-1\Big)+\widehat{\beta}_{t}S_{x}\sum_{j=1}^{m_{i}}\frac{y_{ij}\widehat{\mu}_{i+}}{(\widehat{\phi}+\widehat{\mu}_{i+})^2}.
\end{equation*}

\subsection*{\underline{Dispersion parameter perturbation}}

Let $\phi^{*}_i= \omega_i \times \phi$ be the dispersion parameter perturbation, where $\omega_i \in \mathbb{R}^{+}$. The log-likelihood function of the perturbed model takes the form

\begin{align*}
\ell(\bm{\theta}|\bm{\omega}) &= \sum_{i=1}^{n}\Bigg\{\log\Big( \Gamma(\phi^{*}_i + y_{i+}) \Big) - \log\Big( \Gamma(\phi^{*}_i) \Big) -
                             \sum_{j=1}^{m_i}\log(y_{ij}!) + \phi^{*}_i\log(\phi^{*}_i) \\
                            &\quad{-} \phi^{*}_i\log(\phi^{*}_i + \mu_{i+}) + \sum_{j=1}^{m_i}y_{ij}\log(\mu_{ij}) -
                             \sum_{j=1}^{m_i}y_{ij}\log(\phi^{*}_i + \mu_{i+})\Bigg\}.
\end{align*}
 $\bm{\Delta}=(\bm{\Delta}_{\bm \beta}^{\top}, \bm{\Delta}^{\top}_{\phi})^{\top}$, where the $i$-th elements of $\bm{\Delta}^{\top}_{\bm \beta}$ and $\bm{\Delta}^{\top}_{\phi}$, respectively, are given by
\begin{align*}
\Delta_{k_i}&=\frac{\widehat{\phi}^{*}_{i}\widehat{\mu}_{i+}(y_{i+}-\widehat{\mu}_{i+})}{(\widehat{\phi}^{*}_{i} + \widehat{\mu}_{i+})^2}x_{ijk} ~ ~{\rm and} \\
\Delta_{\phi_i}&=\psi( \widehat{\phi}^{*}_{i} + y_{i+}) - \psi(\widehat{\phi}^{*}_{i}) + \widehat{\phi}^{*}_{i}[\psi^{\prime}(\widehat{\phi}^{*}_{i}+ y_{i+}) - \psi^{\prime}(\widehat{\phi}^{*}_{i})] + 1 \\
                & \quad{+}\log\left(\frac{\widehat{\phi}^{*}_{i}}{\widehat{\phi}^{*}_{i}+\widehat{\mu}_{i+}}\right) + \frac{\widehat{\mu}_{i+}}{ \widehat{\phi}^{*}_{i}+\widehat{\mu}_{i+}}+ \frac{y_{i+} \widehat{\mu}_{i+}}{(\widehat{\phi}^{*}_{i}+\widehat{\mu}_{i+})^2 } +
                  \frac{\widehat{\phi}^{*}_{i}( 2\widehat{\mu}_{i+} + \widehat{\phi}^{*}_{i})}{(\widehat{\phi}^{*}_{i}+\widehat{\mu}_{i+})^2}.
\end{align*}
For all the perturbation schemes, the $\ddot{\ell}(\bm{\theta})$ matrix in $C_{\bm{d}}(\bm{\theta})$ is described  in the Appendix.

\section{Numerical results}
\label{sec:simu}

First, a simulation study is conducted to evaluate the asymptotic behavior of the ML estimator,
$\widehat{\bm{\theta}}$. The vector $\bm{y}=(\bm{y}^{\top}_1,\ldots,\bm{y}^{\top}_n)^{\top},$ 
where $\bm{y}_i\sim \text{MNB}(\bm{\mu}_{i},\phi)$, is generated from a random intercept 
Poisson-GLG model with
$(i)$~$y_{ij} |b_i$ $\stackrel{\rm ind} {\sim}$ ${\rm Poisson}(u_{ij})$, 
$(ii)$~$\log(u_{ij}) = \beta_0 + \beta_1 x_{1ij} + \beta_2 x_{2ij} + b_i$, and  
$(iii)$~$b_{i} \stackrel{\rm iid} {\sim} {\rm GLG}(0, \lambda, \lambda),$ 
where  $x_{1ij} \sim N(0,1)$ and $x_{2ij}$ is a dummy variable with two levels, 
 for $i=1,\ldots,n$ and $j=1,2,3$.  Further, it is assumed that $\bm{\beta}=(1.5,1.0,0.0)^{\top}$
 and $\lambda=\phi^{-1/2}$ in the GLG distribution. The Bias, root mean square error (RMSE) and coverage probabilities for the $95\%$ confidence level are obtained from $R=10,000$ Monte Carlo replications performed for a sample size of $n=50$, $100$, $150$, and $200$ and for three different values of the dispersion parameter, namely, $\phi=3$, $5$ and $7$. The estimates of the Bias and RMSE for $\bm{\theta}$ are obtained from: ${\rm Bias}(\theta_s)=\overline{\theta}_{s}-\theta_{s}$ and ${\rm RMSE}(\theta_{s})=\sqrt{\sum_{r=1}^{R}(\widehat{\theta}^{(r)}_{s}-\theta_{s})^{2}/R}$, respectively, where $\overline{\theta}_{s}=\sum_{r=1}^{R}\widehat{\theta}_{s}^{(r)}/R$ and $\widehat{\theta}^{(r)}_{s}$ the estimate from $s$th parameter obtained in $r$th Monte Carlo replication.
These simulation results are presented in Table \ref{table1}. As expected, the Bias and RMSE values of the regression coefficients estimates decrease as the sample size and values of the $\phi$ parameter increase. The coverage probabilities are close to 0.95. Moreover, the results show that the dispersion parameter exhibits a desirable asymptotic behavior when $\phi$ assumes small values. Since $\phi = \lambda^{-2}$, it is possible to affirm that the asymptotic
properties of the dispersion parameter estimator are associated with the asymmetry  of the GLG distribution.
\begin{table}[h!]
\centering {
\caption{Bias, root mean square error (RMSE), and coverage (\%) of 95\% confidence intervals.}
\label{table1}
\begin{small}
\begin{tabular}{cccrrrr}
\hline
              $\phi (\lambda)$   &      Measure           &  $n$  & $\phi$& $\beta_0$ & $\beta_1$ & $\beta_2$  \\
\hline
\multirow{12}{*}{3.0(0.58)}  & \multirow{4}{*}{Bias}	& 50    & 0.3931 & -0.0078 & -0.0014 & -0.0029    \\
											 & 												& 100 	&	0.1778 & -0.0041 &  0.0003 &  0.0000    \\
											 &												& 150 	&	0.0995 & -0.0034 &  0.0001 &  0.0000     \\
											 &												& 200 	&	0.0735 & -0.0031 & -0.0002 &  0.0000    \\
\cline{2-7}
											 &	\multirow{4}{*}{RMSE}	& 50    & 1.0345 & 0.1298 & 0.0516 & 0.1821    \\
											 &												& 100 	& 0.5951 & 0.0920 & 0.0335 & 0.1280    \\
											 &												& 150 	& 0.4387 & 0.0748 & 0.0256 & 0.1031     \\
											 &												& 200 	& 0.3793 & 0.0642 & 0.0251 & 0.0886    \\
\cline{2-7}
											 &	\multirow{4}{*}{Coverage}	& 50    & 96.72 & 93.92 & 94.61 & 93.73     \\
											 &												& 100 	& 96.00 & 94.33 & 95.17 & 94.23    \\
											 &												& 150 	& 95.75 & 94.80 & 95.15 & 94.73     \\
											 &												& 200 	& 95.28 & 94.93 & 95.22 & 94.93   \\
\hline
\multirow{12}{*}{5.0(0.44)} & \multirow{4}{*}{Bias}	  & 50    & 0.8473 & -0.0063 & 0.0002 & 0.0000   \\
											 &												& 100 	&	0.3264 & -0.0020 &-0.0003 & 0.0000    \\
											 &												& 150 	&	0.2179 & -0.0021 & 0.0001 & 0.0000     \\
											 &												& 200 	&	0.1583 & -0.0009 & 0.0001 & 0.0000   \\
\cline{2-7}
											 & \multirow{4}{*}{RMSE}	& 50    & 2.1448 & 0.1078 & 0.0492 & 0.1505     \\
											 &												& 100 	& 1.1021 & 0.0741 & 0.0299 & 0.1038    \\
											 &												& 150 	& 0.8531 & 0.0613 & 0.0264 & 0.0829     \\
											 &												& 200 	& 0.7190 & 0.0534 & 0.0228 & 0.0724    \\
\cline{2-7}
											 & \multirow{4}{*}{Coverage} 	& 50    & 97.22 & 93.93 & 94.94 & 93.51     \\
											 &												& 100 	& 95.94 & 94.59 & 95.02 & 94.26    \\
											 &												& 150 	& 95.32 & 94.59 & 94.87 & 95.21     \\
											 &												& 200 	& 95.18 & 94.64 & 95.24 & 94.87   \\
\hline
\multirow{12}{*}{7.0(0.38)}  & \multirow{4}{*}{Bias}	& 50    & 1.3985 & -0.0057 & 0.0000 & 0.0000 \\
											 &												& 100 	&	0.5524 & -0.0031 & 0.0000 & 0.0000    \\
											 &												& 150 	&	0.3459 & -0.0019 & 0.0000 & 0.0000     \\
											 &												& 200 	&	0.2485 & -0.0014 & 0.0000 & 0.0000    \\
\cline{2-7}
											 & \multirow{4}{*}{RMSE}	& 50    & 3.6728 & 0.0976 & 0.0470 & 0.1343  \\
											 &												& 100 	& 1.7556 & 0.0668 & 0.0304 & 0.0901   \\
											 &												& 150 	& 1.3277 & 0.0540 & 0.0256 & 0.0731    \\
											 &												& 200 	& 1.1060 & 0.0473 & 0.0219 & 0.0632    \\
\cline{2-7}
											 & \multirow{4}{*}{Coverage} 	& 50    & 97.13 & 93.97 & 95.01 & 93.36     \\
											 &												& 100 	& 96.92 & 94.54 & 94.94 & 94.83    \\
											 &												& 150 	& 95.83 & 94.80 & 94.85 & 95.05     \\
											 &												& 200 	& 95.81 & 94.76 & 94.76 & 95.07   \\
\hline
\end{tabular}
\end{small}
}
\end{table}
A second simulation study is performed to evaluate the impact of misspecifying the random effect distribution
on the ML estimates of the MNBR model. It is assumed that the random intercept is normally distributed and negatively correlated. For sample size $n=50,100,$ and $150$, the vector $\bm{y}=(\bm{y}^{\top}_1,\ldots,\bm{y}^{\top}_n)^{\top}$ is generated from a random intercept Poisson-Normal distribution with the following hierarchical structure: 
$(i)$~$y_{ij} |b_{i}$ $\stackrel{\rm ind} {\sim}$ ${\rm Poisson}(u_{ij})$, $(ii)$~$\log(u_{ij}) = \beta_0 + \beta_1 x_{1ij} + b_{i}$ and $(iii)$~$b_{i} \stackrel{\rm iid} {\sim} {\rm N}(0, \sigma^{2})$, $\sigma^{2}=\phi^{-1}$ and $(i)$~$y_{ij} |b_{ij}$ $\stackrel{\rm ind} {\sim}$ ${\rm Poisson}(u_{ij})$, $(ii)$~$\log(u_{ij}) = \beta_0 + \beta_1 x_{1ij} + z_{1ij}b_{ij}$, and $(iii)$~$\bm{b} {\sim} {\rm N}_{3}(\bm{0}, \bm{\Sigma}),$ where  $(\beta_0,\beta_1)^{\top}=(1.5,1.0)$, $x_{1ij} \sim U(0,1)$, 
\begin{eqnarray*}
\bm{\Sigma} = \Bigg[
		\begin{tabular}{ccc}
			$\phi^{-1}$ & -0.5 & -0.1 \\
		 -0.5 & $\phi^{-1}$ & -1.0 \\
		 -0.1 & -1.0 &  $\phi^{-1}$ \\
		\end{tabular}
\Bigg],
\end{eqnarray*}
and $\phi^{-1}=4.0$. In Table \ref{table4}, we present the Bias and root mean square error (RMSE),  calculated by fitting the MNBR model on 1,000 Monte Carlo replications
for different sample sizes. Based on the assumption that $b_{i} \stackrel{\rm iid} {\sim} {\rm N}(0, \sigma^{2})$, we 
observe that the parameter estimates of $\phi$ and $\beta_0$ are biased.  
To summarize, we conclude that the MNBR model provides inconsistent estimates of the asymptotic variance of the ML estimators when the covariance matrix of the random effect is misspecified.
\begin{table}[h!]
\centering {
\caption{Bias and root mean square error (RMSE) when the covariance matrix of the random response variable is misspecified.}
\label{table4}
\begin{small}
\begin{tabular}{cccrrr}
\hline
 Variance & $n$ & Measure & $\phi$ & $\beta_0$ & $\beta_1$  \\
\hline
\multirow{6}{*}{$\sigma^{2}$}
& 	\multirow{2}{*}{50}									  & Bias 	& -0.4285 & -1.8818 &  0.0561   \\
& 											& RMSE 	&  0.4584 &  1.9805 &  1.0931   \\
\cline{2-6}
& 	\multirow{2}{*}{100}										& Bias 	& -0.3799 & -1.9408 & 0.0497   \\
& 											& RMSE 	&  0.3968 &  2.0068 & 0.9257   \\
\cline{2-6}
& 	\multirow{2}{*}{150}										& Bias 	& -0.3633 & -1.9282 & -0.0129   \\
& 											& RMSE 	&  0.3750 &  1.9754 &  0.6798   \\
\hline
\multirow{6}{*}{$\bm{\Sigma} $}
& 	\multirow{2}{*}{50}								  & Bias 	& -0.4965 & -1.8623 & 0.0068  \\
& 											& RMSE 	&  0.5301 &  1.9669 & 1.0656  \\
\cline{2-6}
& 	\multirow{2}{*}{100}										& Bias 	& -0.4375 & -1.9014 & -0.0315   \\
& 											& RMSE 	&  0.4560 &  1.9601 &  0.8784   \\
\cline{2-6}
& 	\multirow{2}{*}{150}										& Bias 	& -0.4196 & -1.9228 & -0.0372   \\
& 											& RMSE 	&  0.4353 &  1.9953 &  0.8282   \\
\hline
\end{tabular}
\end{small}
}
\end{table}

A third simulation study is conducted to evaluate the variance-to-mean ratio, ${\rm VMR}={\rm Var}(\bm{y})/{\rm E}(\bm{y})$. Considering three different values for the dispersion parameter, $\phi =0.1, 1.0$ and $50$, random samples are generated and the VMR statistic are computed. The ranges to VMR, for each value of $\phi$, are $(122.39,220.24)$, $(19.38,29.35)$ and $(5.85,16.87)$, respectively. In the Table \ref{table_ratio}, we reported the bias, RMSE and coverage of the confidence intervals. As expected, Table \ref{table_ratio} shows that the ML estimates present a desirable asymptotic behavior when the sample size increases. In addition, we observe that when $\phi$ increases the VMR statistic decreases and assumes values more than one. Thus, it is possible to conclude that the MNBR model is an overdispersion model. When the dispersion parameter is more than 50 similar results are obtained.

\begin{table}[h!]
\centering {
\caption{Bias, root mean square error (RMSE), coverage (\%) of 95\% confidence intervals.}
\label{table_ratio}
\begin{small}
\begin{tabular}{cccrrrrc}
\hline
    $\phi$   &      Measure           &  $n$  & $\phi$& $\beta_0$ & $\beta_1$ & $\beta_2$ & VMR \\
\hline
\multirow{12}{*}{0.1}  
                       & \multirow{4}{*}{Bias}	      & 50    & 0.0093 & -0.1769 & -0.0006 &  0.0575  & \multirow{12}{*}{(122.39,220.24)} \\
											 & 												      & 100 	& 0.0051 & -0.1201 & -0.0004 &  0.0347  &  \\
											 &												      & 150 	& 0.0027 & -0.0838 & -0.0003 &  0.0310  &  \\
											 &												      & 200 	& 0.0020 & -0.0331 & -0.0001 &  0.0226  &  \\
\cline{2-7}
											 &	\multirow{4}{*}{RMSE}	      & 50    & 0.0299 & 0.7048 & 0.0707 & 0.9794  &  \\
											 &												      & 100 	& 0.0198 & 0.5105 & 0.0411 & 0.6742  &  \\
											 &												      & 150 	& 0.0151 & 0.3891 & 0.0298 & 0.5531  &   \\
											 &												      & 200 	& 0.0130 & 0.3203 & 0.0253 & 0.4468  &  \\
\cline{2-7}
											 &	\multirow{4}{*}{Coverage}	  & 50    & 95.4 & 91.3 & 95.0 & 91.1  &  \\
											 &												      & 100 	& 95.0 & 92.8 & 94.2 & 92.8  &  \\
											 &												      & 150 	& 95.3 & 93.7 & 94.9 & 92.9  &  \\
											 &												      & 200 	& 95.5 & 94.5 & 95.1 & 94.7  &  \\
\hline
\multirow{12}{*}{1.0}  & \multirow{4}{*}{Bias}	      & 50    & 0.0979 & -0.0201 & 0.0017 & -0.0026 & \multirow{12}{*}{(19.38,29.35)}  \\
											 & 												      & 100 	& 0.0390 & -0.0059 & 0.0000 & -0.0189 \\
											 &												      & 150 	& 0.0317 & -0.0045 & 0.0000 & -0.0160 \\
											 &												      & 200 	& 0.0164 & -0.0039 & 0.0000 & -0.0006 \\
\cline{2-7}
											 &	\multirow{4}{*}{RMSE}	      & 50    & 0.2706 & 0.2188 & 0.0552 & 0.3046 \\
											 &												      & 100 	& 0.1627 & 0.1550 & 0.0380 & 0.2132  \\
											 &												      & 150 	& 0.1351 & 0.1249 & 0.0291 & 0.1712 \\
											 &												      & 200 	& 0.1065 & 0.1069 & 0.0244 & 0.1477 \\
\cline{2-7}
											 &	\multirow{4}{*}{Coverage}	  & 50    & 95.7 & 93.9 & 94.7 & 92.2 \\
											 &												      & 100 	& 95.4 & 94.3 & 95.6 & 93.5 \\
											 &												      & 150 	& 95.3 & 94.9 & 95.2 & 94.1\\
											 &												      & 200 	& 95.1 & 95.2 & 94.9 & 94.7 \\
\hline
\multirow{12}{*}{50}     & \multirow{4}{*}{Bias}	      & 50    & 36.1329 & -0.0034 & 0.0000 &  0.0022 & \multirow{12}{*}{(5.85,16.87)}  \\
											 & 												      & 100 	& 17.2463 & -0.0030 & 0.0000 &  0.0015  & \\
											 &												      & 150 	& 11.0381 & -0.0009 & 0.0000 &  0.0011  & \\
											 &												      & 200 	&  9.1020 & -0.0008 & 0.0000 &  0.0021  & \\
\cline{2-7}
											 &	\multirow{4}{*}{RMSE}	      & 50    &  73.4328 & 0.0621 & 0.0332 & 0.0739 &  \\
											 &												      & 100 	&  45.1901 & 0.0433 & 0.0272 & 0.0519 &  \\
											 &												      & 150 	&  30.6868 & 0.0367 & 0.0213 & 0.0454 & \\
											 &												      & 200 	&  25.7042 & 0.0312 & 0.0206 & 0.0374 & \\
\cline{2-7}
											 &	\multirow{4}{*}{Coverage}	  & 50    &  95.0 & 95.0 & 93.2 & 95.3 & \\
											 &												      & 100 	&  94.2 & 95.5 & 94.1 & 95.4 &  \\
											 &												      & 150 	&  94.0 & 93.9 & 94.0 & 94.1 & \\
											 &												      & 200 	&  95.7 & 93.4 & 94.1 & 94.5 & \\
\hline
\end{tabular}
\end{small}
}
\end{table}
\section{Data analysis}
\label{apli}

In this section, we present two examples illustrating the diagnostic tools for the MNBR model. The data sets and programs can be found in the \texttt{MNB} package available on the \texttt{R} platform. 

\subsubsection*{\underline{Seizures data}}

\begin{figure}
\begin{minipage}[b]{0.32\linewidth}
\includegraphics[width=\linewidth]{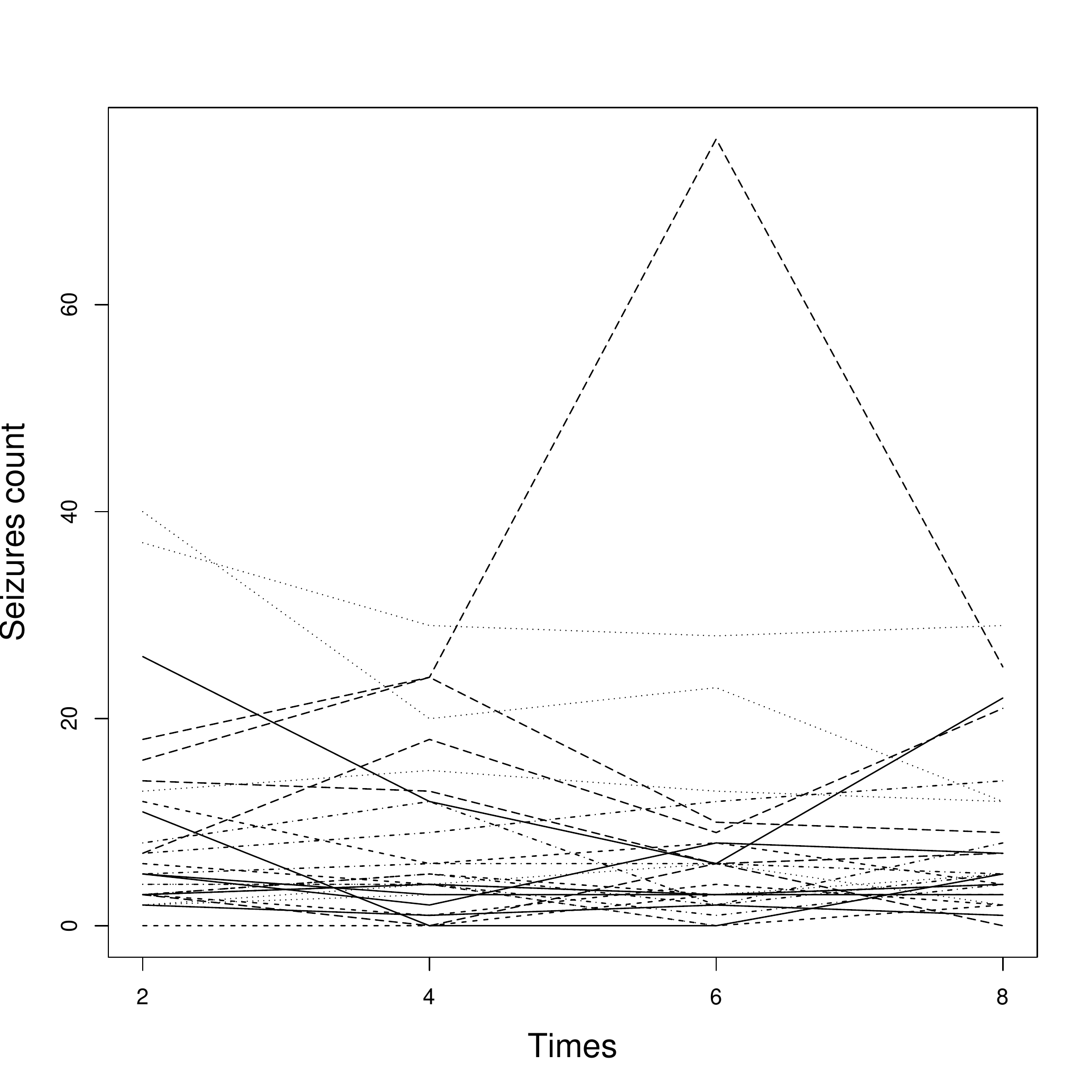}
\begin{center}
(a)
\end{center}
\end{minipage}
\hfill
\begin{minipage}[b]{0.32\linewidth}
\includegraphics[width=\linewidth]{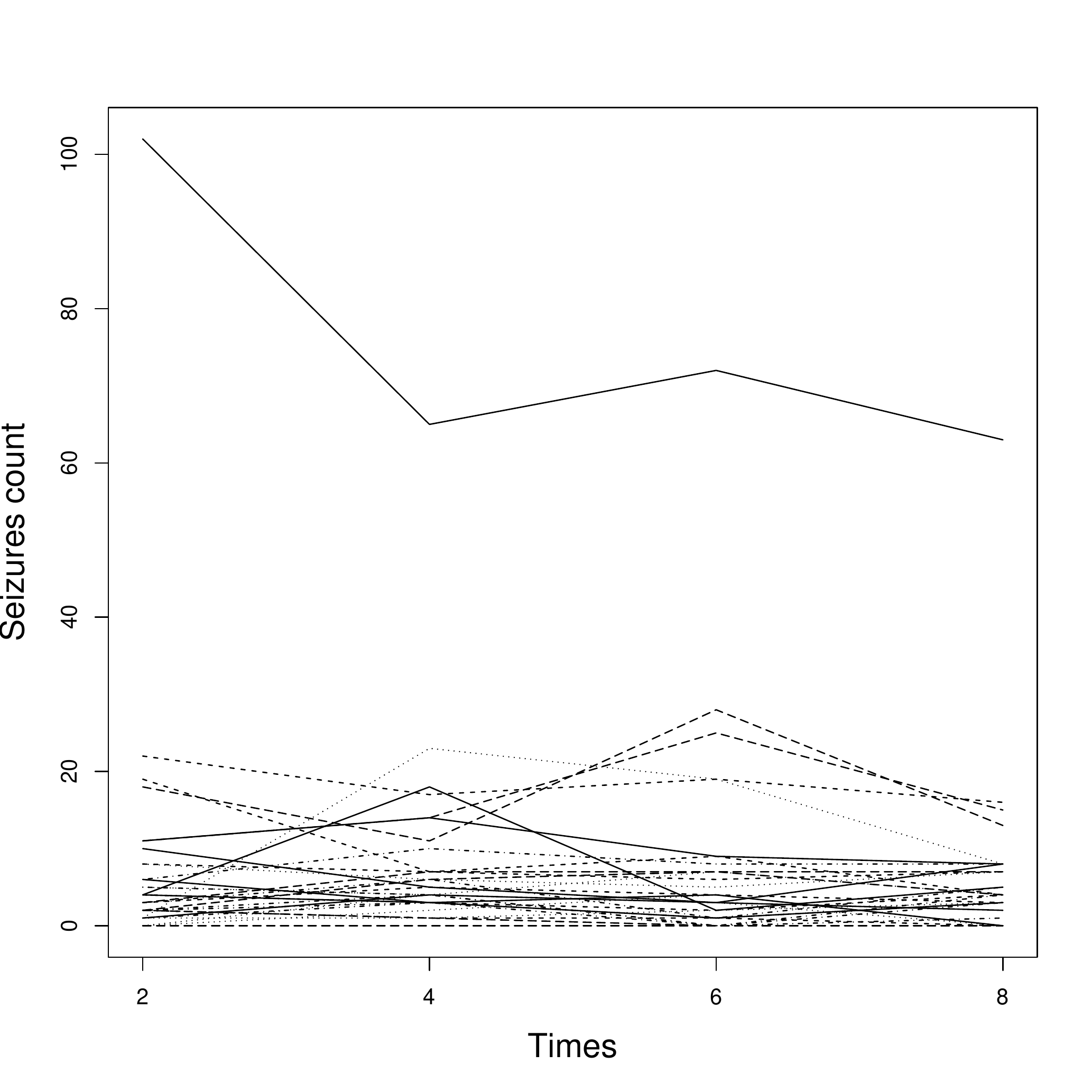}
\begin{center}
(b)
\end{center}
\end{minipage}
\begin{minipage}[b]{0.32\linewidth}
\includegraphics[width=\linewidth]{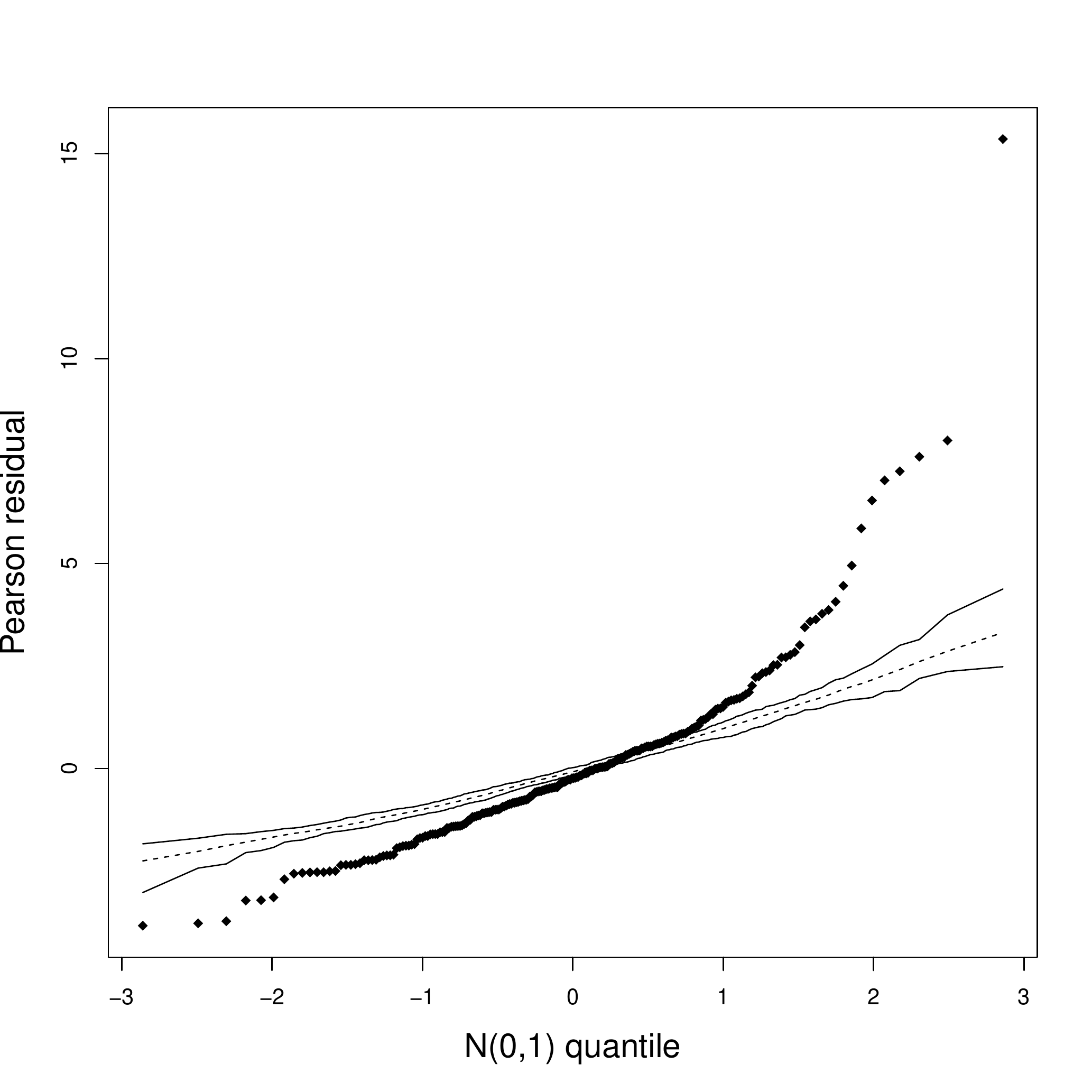}
\begin{center}
(c)
\end{center}
\end{minipage}
\caption{Individual profiles of the patients in (a) the placebo group, (b) the progabide group, and (c) the simulated envelope plot of the Pearson residuals.}
\label{resid}
\end{figure}

\begin{figure}[h!]
\begin{minipage}[b]{0.48\linewidth}
\includegraphics[width=\linewidth]{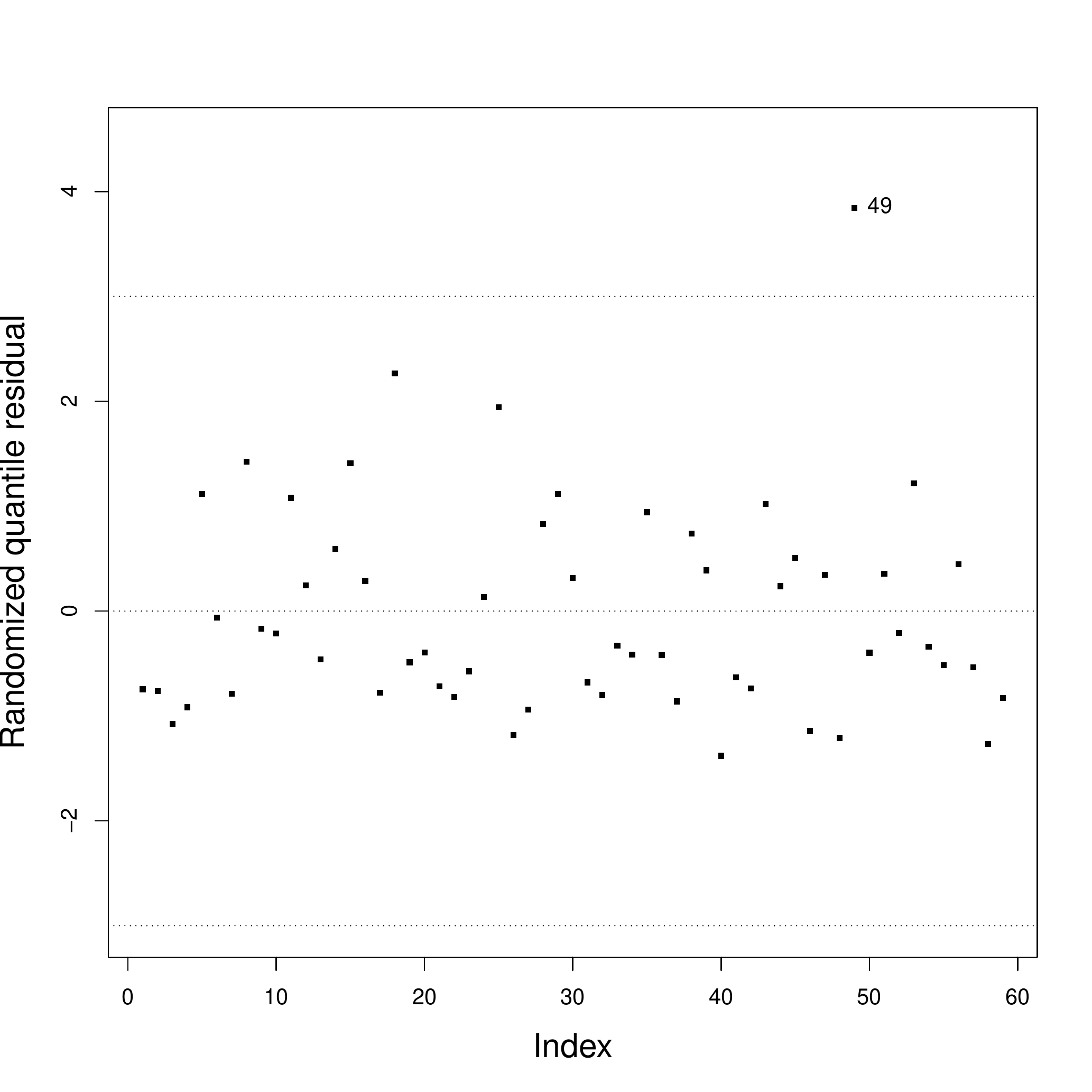}
\begin{center}
(a)
\end{center}
\end{minipage}
\hfill
\begin{minipage}[b]{0.48\linewidth}
\includegraphics[width=\linewidth]{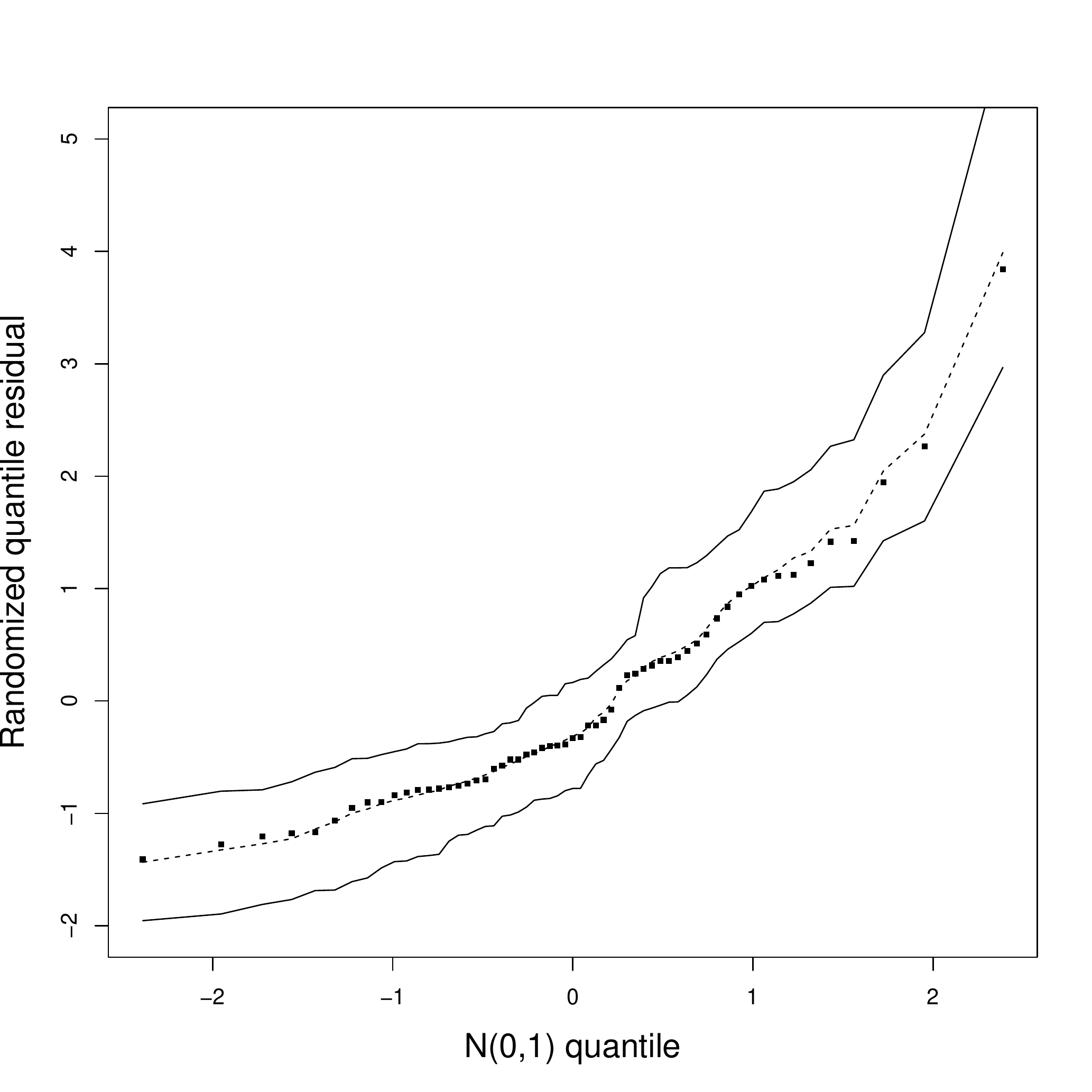}
\begin{center}
(b)
\end{center}
\end{minipage}
\caption{Index plot of the randomized quantile residuals (a) and (b) the simulated envelope plot of the randomized quantile residuals.}
\label{resi}
\end{figure}

The data set described in \cite{Diggle04} refers to an experiment in which 59 epileptic patients were randomly assigned to one of two
treatment groups: treatment (progabide drug) and placebo groups. The number of seizures experienced by each patient
during the baseline period (week eight) and the four consecutive periods (every two weeks) was recorded. The main goal
of this application is to analyze the drug effect with respect to the placebo. Two dummy covariates are considered in this study; Group which assumes values equal to 1 if the patient belongs to treatment group and 0 otherwise, and Period which assumes values equal to 1 if the number of seizures are recorded during the treatment and 0 if are measured in the baseline period. 
Taking into account the irregular measurement of rate seizures during the time, the variable Time is considered as an offset for fitting the data, where Time assumes values equal 8 if the number of seizures is observed in the baseline period and 2 otherwise. The individual profiles of the patients belonging to the placebo and progabide groups are shown in Figure~\ref{resid} (a) and (b), respectively. The atypical individual profiles correspond to patient $\sharp 25 (18,24,76,25)$ of the placebo group, who presented a high number of seizures in the third visit compared to other clinic visits, and patient $\sharp 49 (102,65,72,63)$ of the progabide group, who suffered a high number of seizures in every clinic visit, indicating the ineffectiveness of the drug in patients with complex seizures. In both groups, it is possible to see the right skewness of the empirical distributions of individual profiles.
Figure \ref{resid}(c) shows a normal probability plot with
simulated envelope for the Pearson residual \citep[p.135]{Faraway2016} computed by fitting the Poisson regression model to seizures data. We confirm the occurrence of the overdispersion phenomenon. Thus, based on Figure~\ref{resid}(a) - (c), which shows the asymmetric behavior of the empirical distribution of individual profiles, the MNBR model is proposed for modeling this behavior according to the following structure: 
$(i)$ $\bm{y}_{i} \stackrel{\rm ind} {\sim} {\rm MNB}(\bm{\mu}_{i}, \phi)$ and
$(ii)$ $\log(\mu_{ij}) = \beta_0 + \beta_1 {\rm Group}_{i}  + \beta_2 {\rm Period}_{ij}  + \beta_3 ({\rm Group}_{i} \times {\rm Period}_{ij}) + \log({\rm Time}_{ij}),$
where $\bm{y}_{i}=(y_{i1},y_{i2},y_{i3},y_{i4})^{\top},$ $\bm{\mu}_{i}=(\mu_{i1},\mu_{i2},\mu_{i3},\mu_{i4})^{\top},$ for $i=1,\ldots,59$, $\beta_1$ is the logarithm of the ratio of the average rate of the treatment group to the placebo group at baseline, $\beta_2$ is the logarithm of the ratio of the seizure mean after treatment period to before treatment period for the placebo group, and $\exp(\beta_3)$ is the treatment effect, and it is the ratio of post- to pre-treatment mean seizure ratios between treatment and placebo groups. The parameter estimates obtained by using the \texttt{fit.MNB} function from the \texttt{MNB} package are shown in Table~\ref{Apli1}. It is not observed evidence of the treatment effect. The shape parameter estimate indicates that the variability of individual profiles with respect their average is asymmetric to the left with dispersion equal to 1.607 ($\lambda=1/\sqrt{\phi}=0.789$).
\begin{table}[h!]
\centering {
\caption{Parameter estimates with their respective approximate standard errors (Std. error), z-values, and $p$-values for the MNBR model fitted to the epileptic data.}
\label{Apli1}
\begin{small}
\begin{tabular}{c|rrrr}
\hline
Parameter & Estimate & Std. error & z-value & $p$-value  \\
\hline
$\phi$      &   1.607   &      0.278   &      $-$   &    $-$     \\
$\beta_{0}$ &   1.348   &      0.153   &      8.813   &    $<0.001$     \\
$\beta_{1}$ &   0.028   &      0.211   &      0.131   &    0.896     \\
$\beta_{2}$ &   0.112   &      0.047   &      2.386   &    0.017    \\
$\beta_{3}$ &  -0.105   &      0.065   &     -1.610   &    0.107     \\
\hline
$\lambda$   &   0.789   &              &              &              \\
\hline
\end{tabular}
\end{small}
}
\end{table}
The randomized quantile residual presented in Section \ref{subsec:residuos} is used to investigate the presence
of outliers or any indication of lack of fit. Figure~\ref{resi}(a) and (b) shows the absence of extra variability and evidence that patient $\sharp 49$ is atypical. The \texttt{qMNB} and \texttt{envelope.MNB} functions from the \texttt{MNB} package is used to display the graphics of the randomized quantile residual.

\begin{figure}[h!]
\begin{minipage}[b]{0.48\linewidth}
\includegraphics[width=\linewidth]{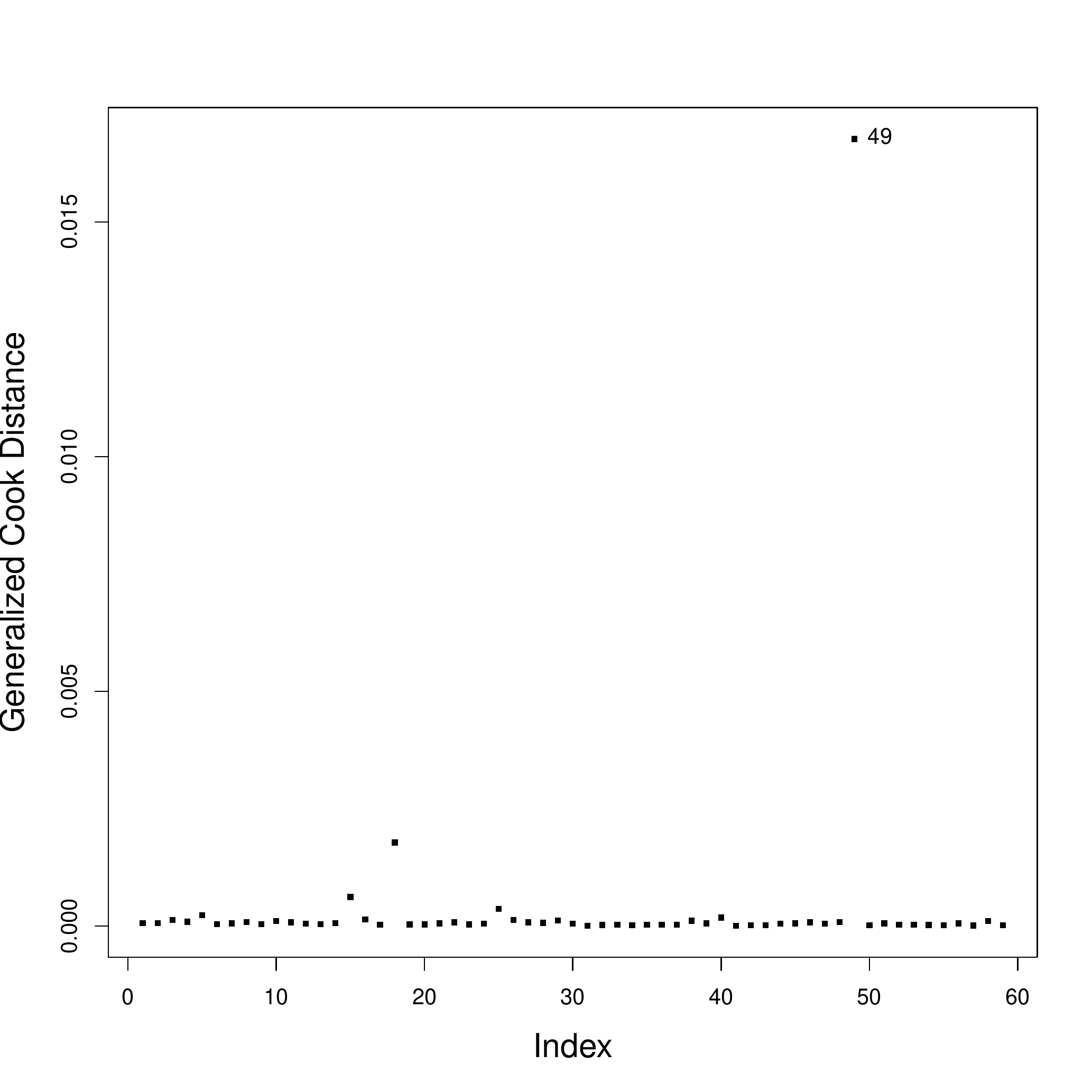}
\begin{center}
(a)
\end{center}
\end{minipage}
\begin{minipage}[b]{0.48\linewidth}
\includegraphics[width=\linewidth]{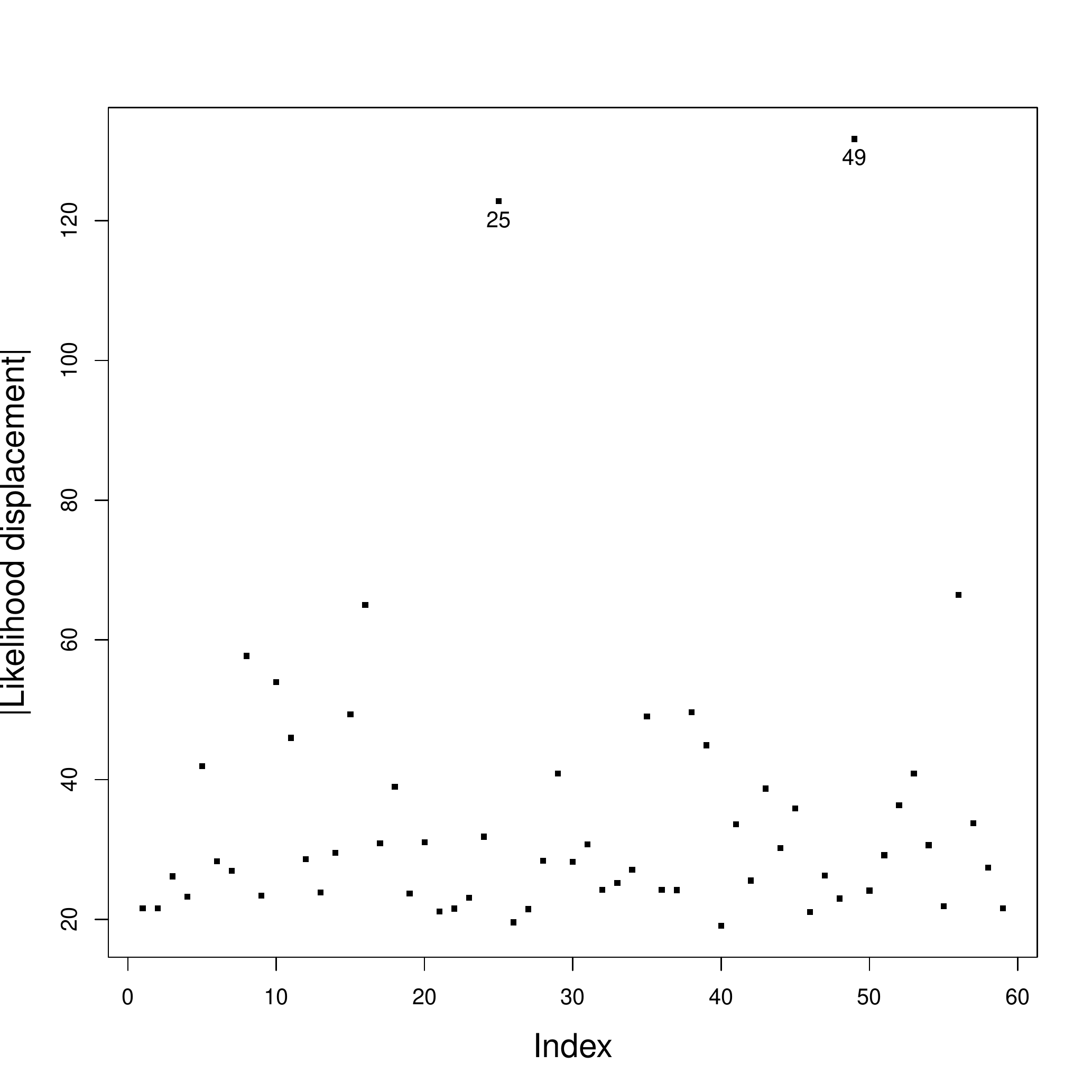}
\begin{center}
(b)
\end{center}
\end{minipage}
\caption{Index plot of (a) the generalized Cook distance and (b) the likelihood displacement.}
\label{global}
\end{figure}

\begin{figure}[h!]
\begin{minipage}[b]{0.48\linewidth}
\includegraphics[width=\linewidth]{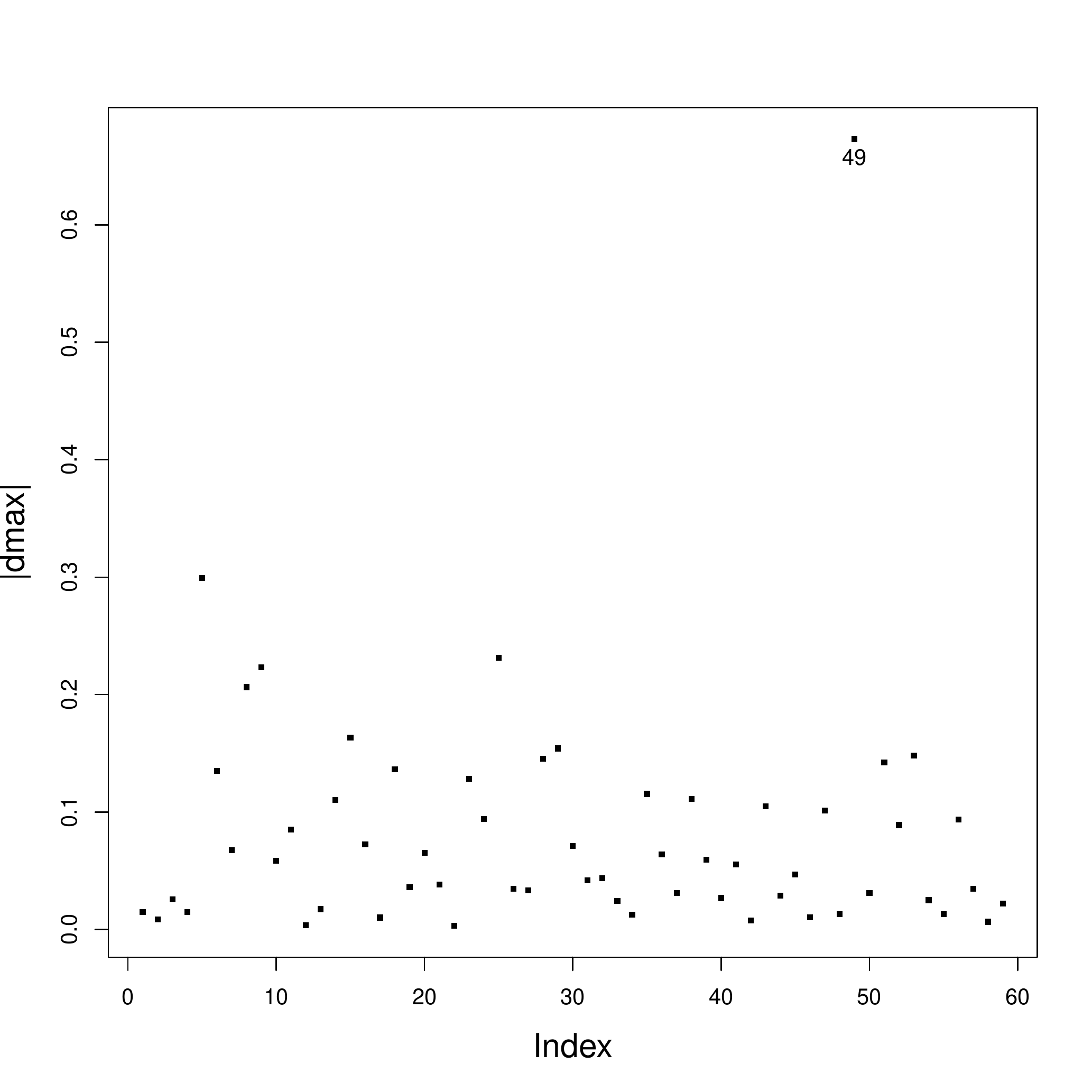}
\begin{center}
(a)
\end{center}
\end{minipage}
\begin{minipage}[b]{0.48\linewidth}
\includegraphics[width=\linewidth]{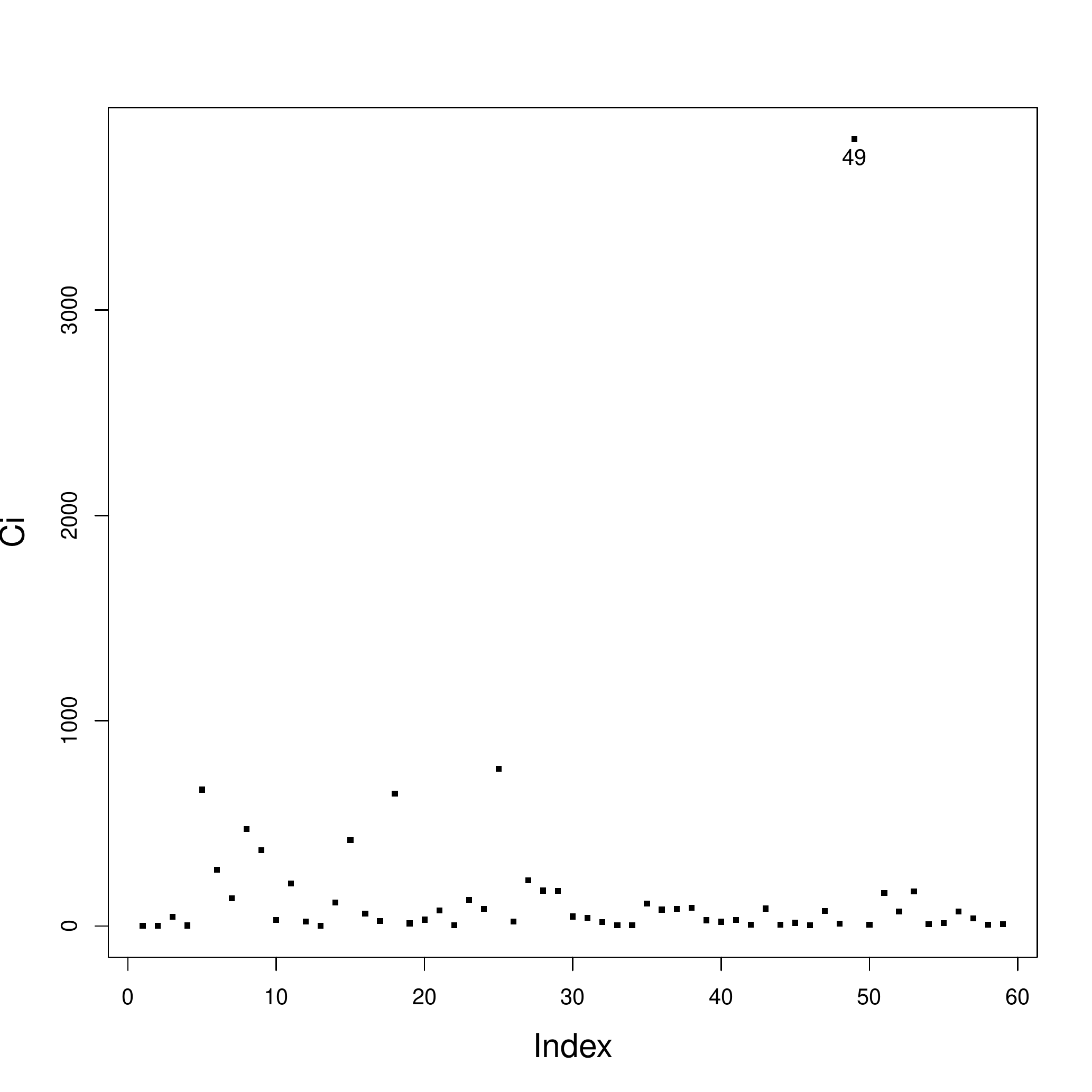}
\begin{center}
(b)
\end{center}
\end{minipage}
\caption{Index plot of $|\bm{d}_{max}|$ for the case weight perturbation scheme: (a) local influence and (b) total local influence on the $i$th patient.}
\label{localseiz}
\end{figure}

\begin{figure}[h!]
\begin{minipage}[b]{0.48\linewidth}
\includegraphics[width=\linewidth]{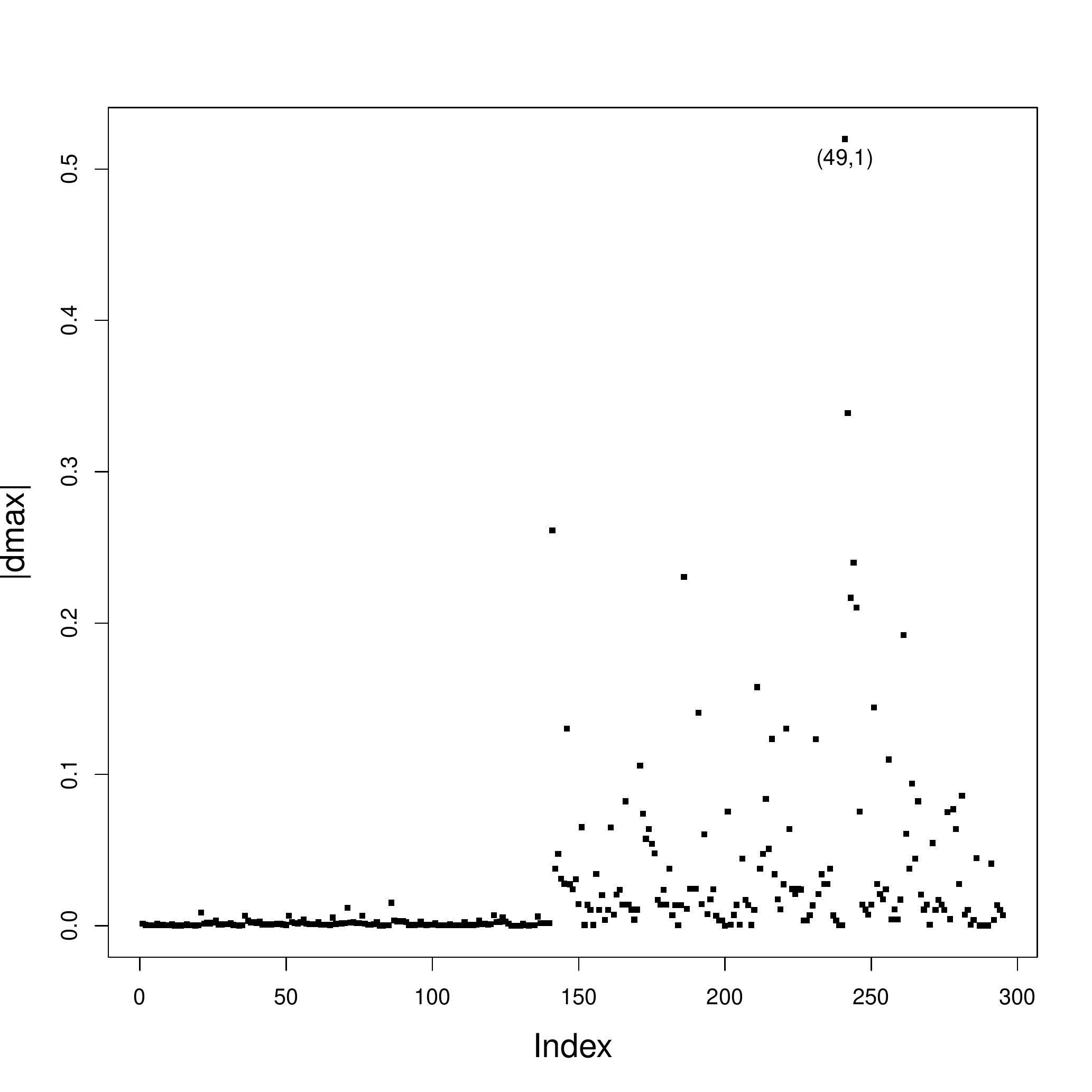}
\begin{center}
(a)
\end{center}
\end{minipage}
\begin{minipage}[b]{0.48\linewidth}
\includegraphics[width=\linewidth]{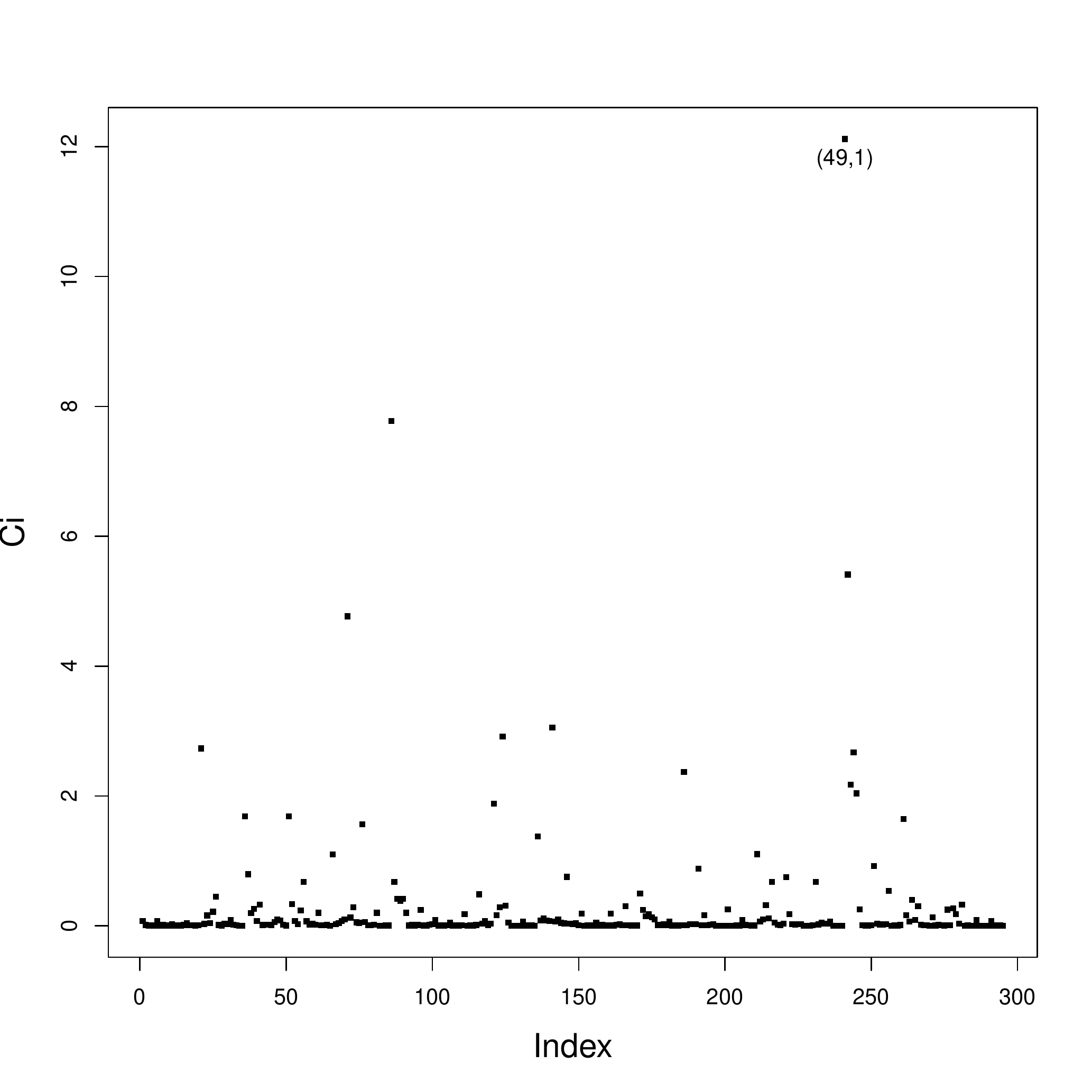}
\begin{center}
(b)
\end{center}
\end{minipage}
\caption{Index plot of $|\bm{d}_{max}|$ for case weight perturbation scheme:  (a) local influence and (b) total influence of the $j$th measurement of each patient.}
\label{totalseiz}
\end{figure}

The global influential graphics presented in Figure \ref{global}(a) and (b) reveal that patients $\sharp 49$, and $\sharp 25$ can impact the ML estimates of the MNBR model when they are removed from the model. The local and total local influential graphics exhibited in Figure \ref{localseiz}(a) and (b) indicate the sensitivity of the estimate associated with patient $\sharp 49$ when the minor perturbation  is induced in the directions of $\bm{d}_{max}$  and $\bm{d}_i,$ respectively. According to
Figure \ref{totalseiz}(a) and (b), the number of epileptic seizures of patient 
$\sharp 49$ recorded on the first and last clinic visits of the trial are  more influential. The dispersion perturbation scheme did not show evidence of any possible influential observations.

Finally, the percentage relative deviations $PRD$=$[(\widehat{\theta}-\widehat{\theta}^{*})/\widehat{\theta}] \times 100 \%$, where $\widehat{\theta}^{*}$ is the estimator of $\theta$ obtained after deleting one or more atypical subjects, are calculated and the results are presented in 
Table~\ref{tabelaDel}.  We can observe significatively changes are associate with the estimates of $\beta_1$ and $\beta_3$. Noting that the signal of the estimate $\beta_1$ is indicating that the treatment decreases the number of seizures, its descriptive-level keeps nonsignificant though. For the parameter $\beta_3$ we observe that the interaction effect incresed 50\% and that its descriptive-level becomes significant.   
\begin{table}[h!]
\centering
{
\caption{Parameter estimates, standard errors (Std. error), $p$-values, and percentage relative deviations ($PRD$).}
\label{tabelaDel}
\begin{small}
\begin{tabular}{c|c|rcrr}
\hline
Dropping  & Parameter & \multicolumn{1}{|c}{Estimate} & \multicolumn{1}{c}{Std. error} & \multicolumn{1}{c}{$p$-value} & \multicolumn{1}{c}{$PRD($\%$)$} \\
\hline
            &  $\phi$      &   2.060   &      0.371   &    $-$   &  -28.21    \\
            &  $\beta_{0}$ &   1.348   &      0.136   &  $<0.000$   &    0.00   \\
$\sharp 49$ &  $\beta_{1}$ &  -0.107   &      0.189   &  0.573   &  487.95   \\
            &  $\beta_{2}$ &   0.112   &      0.047   &  0.017   &    0.00  \\
						&  $\beta_{3}$ &  -0.302   &      0.070   &  $<0.000$   & -188.73	\\	
\hline
\end{tabular}
\end{small}
}
\end{table}

\subsubsection*{\underline{Accident data}}
\begin{figure}[h!]
\begin{minipage}[b]{0.32\linewidth}
\includegraphics[width=\linewidth]{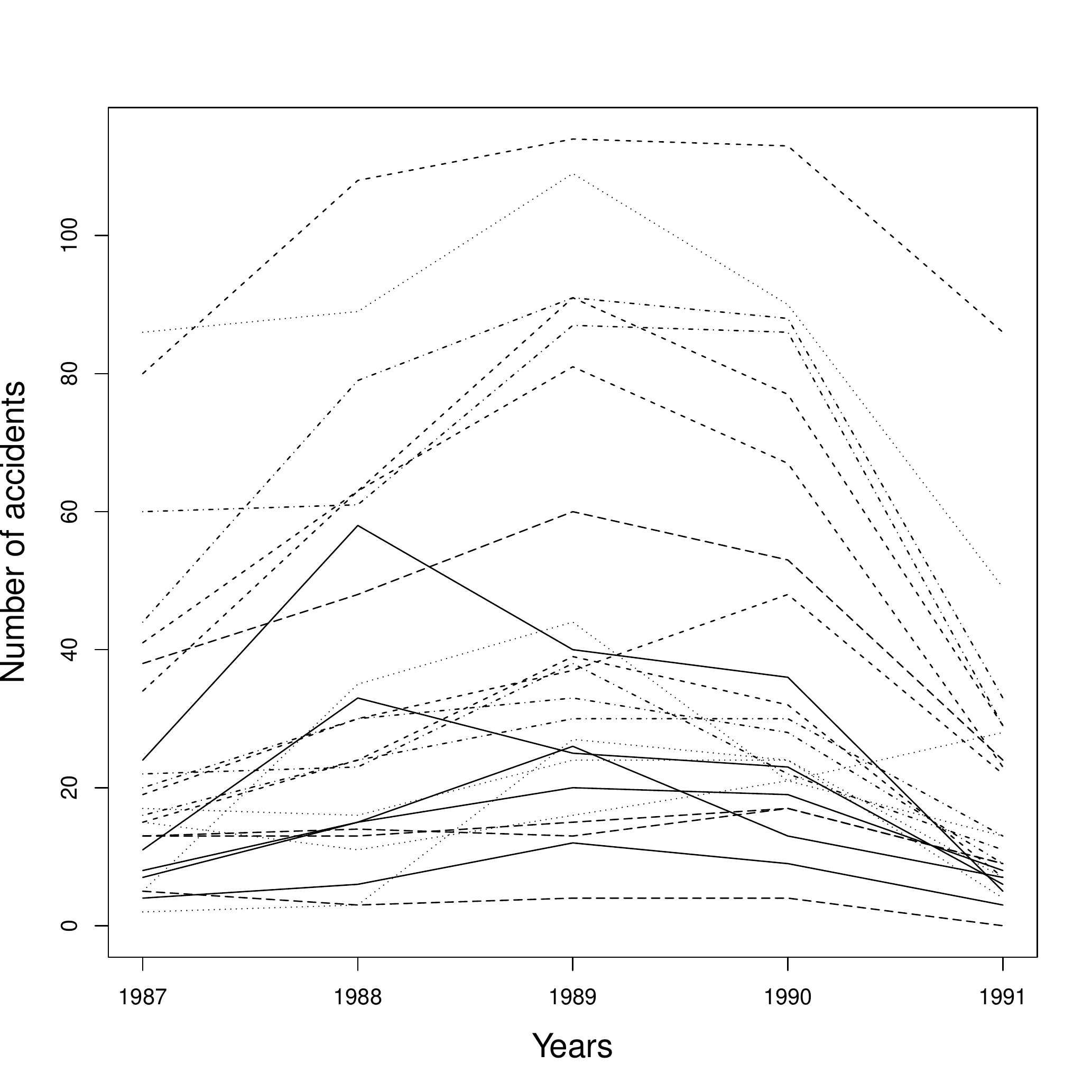}
\begin{center}
(a)
\end{center}
\end{minipage}
\hfill
\begin{minipage}[b]{0.32\linewidth}
\includegraphics[width=\linewidth]{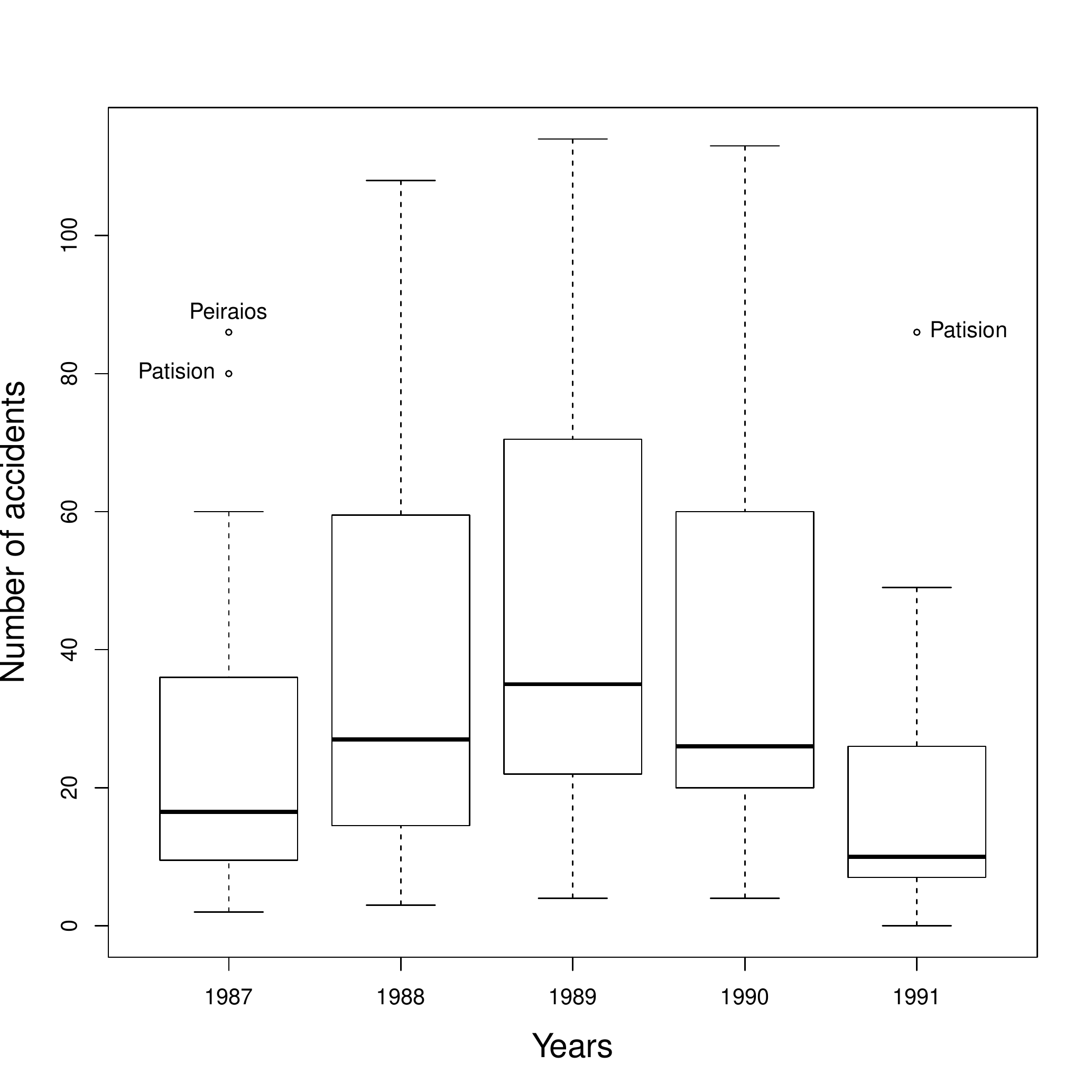}
\begin{center}
(b)
\end{center}
\end{minipage}
\begin{minipage}[b]{0.32\linewidth}
\includegraphics[width=\linewidth]{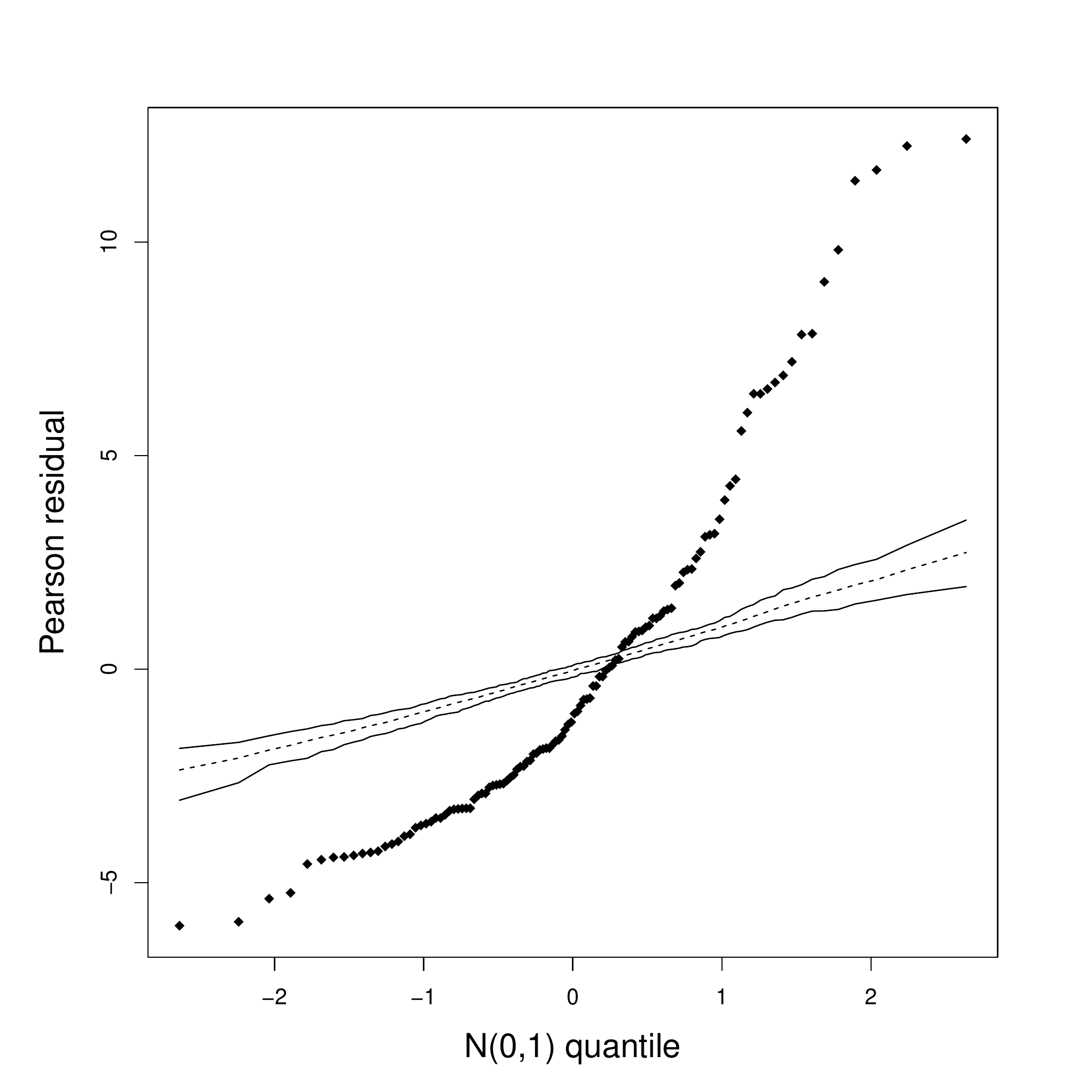}
\begin{center}
(c)
\end{center}
\end{minipage}
\caption{(a) Individual profiles, (b) box-plots of the number of car accidents on  central Athens roads during the 5 years, and (c) simulated envelope plot of Pearson residuals.}
\label{perfi_box}
\end{figure}

\citet{Karlis} provides a data set which refers to the number of car accidents in 24 central roads in Athens, for a time period of 1987 to 1991. 
In this paper, we are interested in modeling the number of car accidents per road length during the period of 5 years. Thus, the dummy covariate years is considered for fitting the data including the road length as an offset. Figure \ref{perfi_box}(a) presents the individual profiles of the central roads for the number of car accidents recorded on each road over time. In general, the highest number of car accidents occurred in 1989 and the smallest in 1991.  
The empirical distribution of individual profiles is skewed to the right. Figure \ref{perfi_box}(b) reveals that $Patision$ and $Peiraios$ roads are considered as outliers and evidence of variability  between the individual profiles and across the years is suggested in Figure~\ref{perfi_box}(c). 

In the fitting the Poisson regression model to the accidents data, Figure \ref{perfi_box}(c), we observe the occurrence of the overdispersion phenomenon. Based on Figure \ref{perfi_box}(a) - (b) showing the asymmetric behavior of the empirical distribution of individual profiles, we propose the MNBR model with the following structure for modeling the data: 
$(i)$ $\bm{y}_{i} \stackrel{\rm ind} {\sim} {\rm MNB}(\bm{\mu}_{i}, \phi)$ and 
$(ii)$ $\log(\mu_{ij})= \beta_{0} + \beta_{t}{\rm year}_{j} + \log(\text{length}_{i})$,
where  
$\bm{y}_i =(y_{i1},y_{i2},y_{i3},y_{i4},y_{i5})^{\top}$, $\bm{\mu}_{i}=(\mu_{i1},\mu_{i2},\mu_{i3},\mu_{i4},\mu_{i5})^{\top}$,
$t=1,2,3,4$, $\beta_t$ is the logarithm of the ratio of the average car accidents per road length for year $t$ to year 1987.
The ML estimates for the MNBR model are presented in Table \ref{esti-Apli2}. We observe the estimates are significant and that the small difference in the logarithm of the rate average of the number of car accidents per length of the roads occurred between the years the 1987 and 1991. 

\begin{figure}[h!]
\begin{minipage}[b]{0.48\linewidth}
\includegraphics[width=\linewidth]{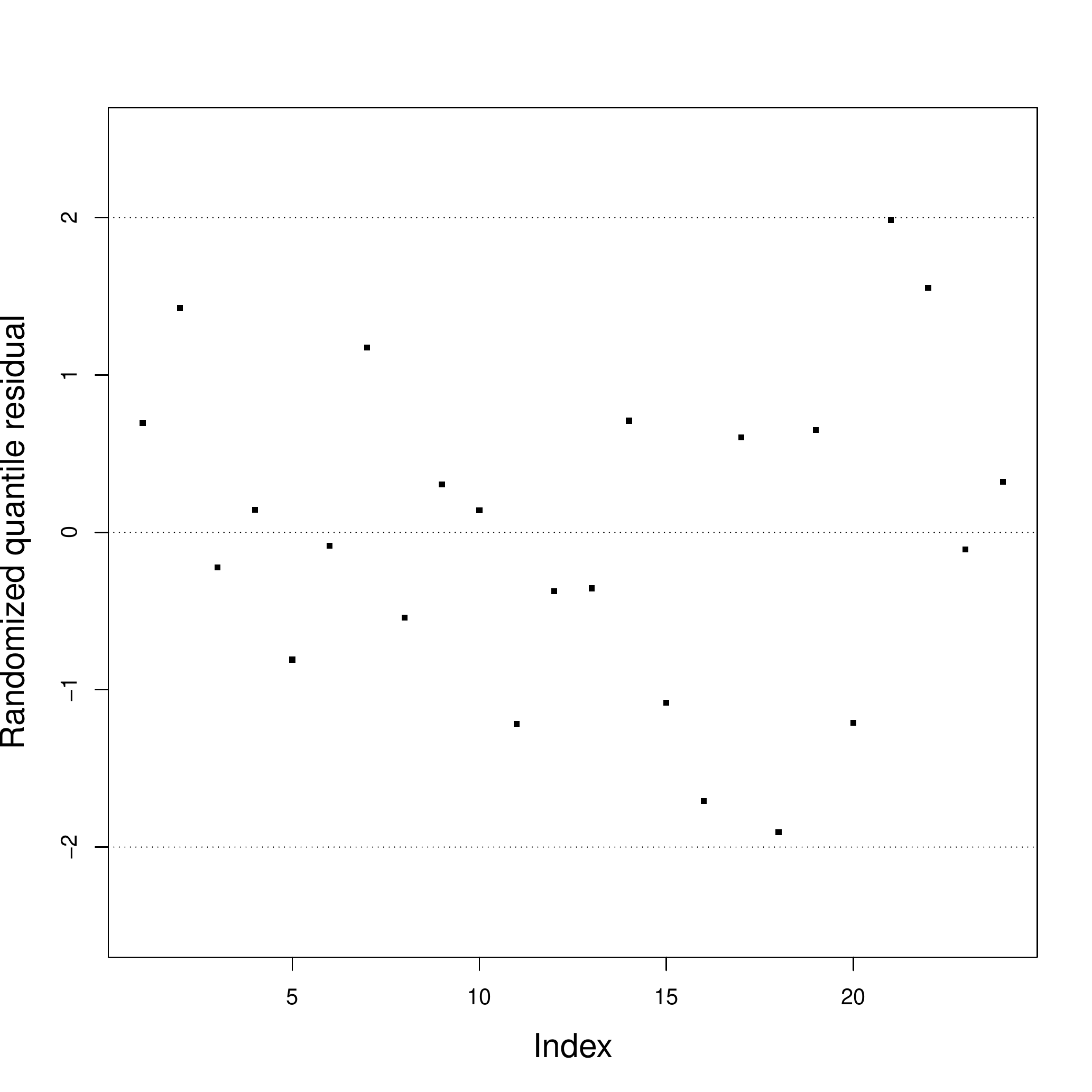}
\begin{center}
(a)
\end{center}
\end{minipage}
\hfill
\begin{minipage}[b]{0.48\linewidth}
\includegraphics[width=\linewidth]{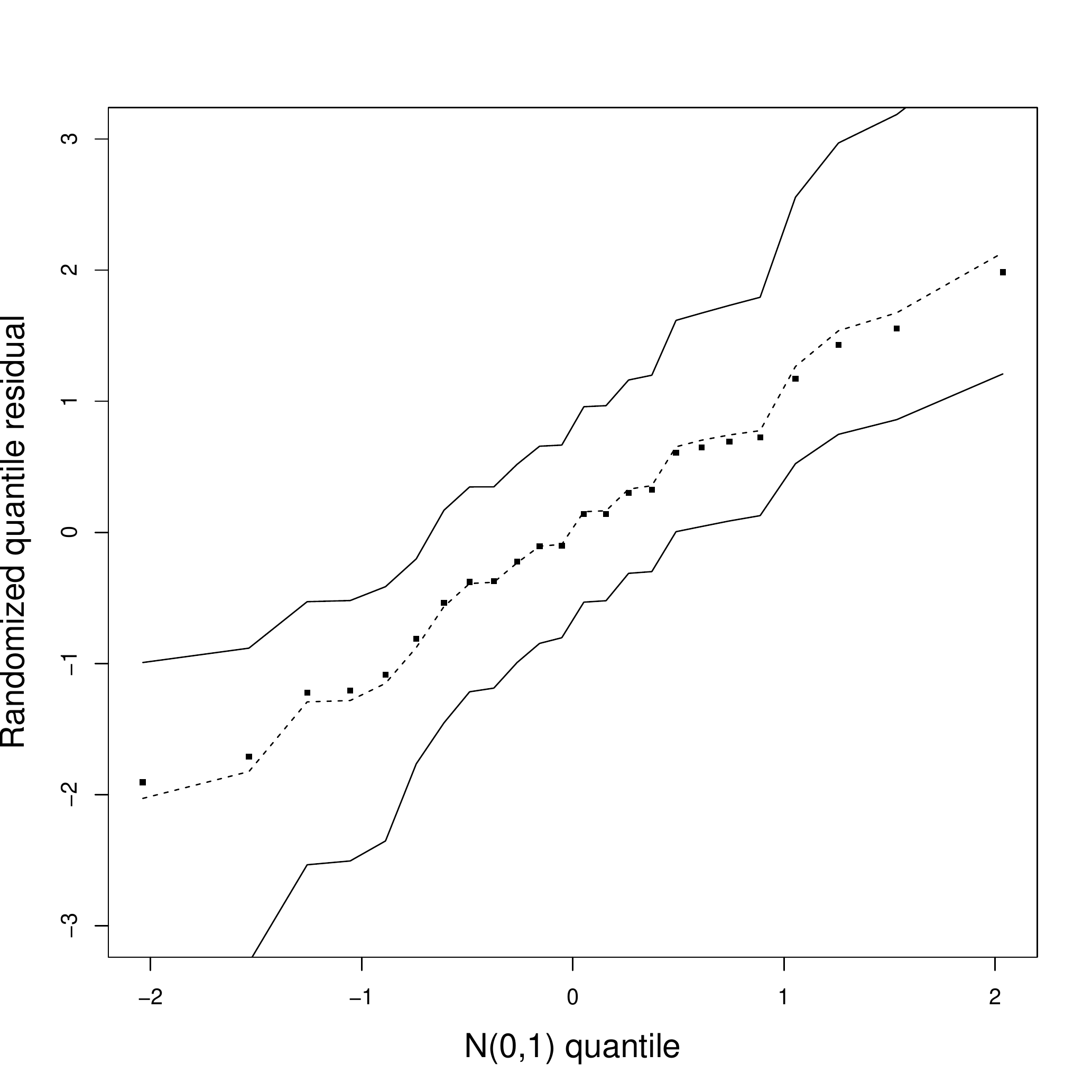}
\begin{center}
(b)
\end{center}
\end{minipage}
\caption{Index plot of the randomized quantile residuals (a) and (b) the simulated envelope plot of the randomized quantile residuals.}
\label{Apli2-resi}
\end{figure}

\begin{table}[h!]
\centering {
\caption{Parameter estimates, standard errors (Std. error), z-values, and $p$-values for the MNBR model fitted to the accidents data.}
\label{esti-Apli2}
\begin{tabular}{c|rrrr}
\hline
Parameter & Estimate & Std. error & z-value & $p$-value  \\
\hline
$\phi$      &  3.364 & 0.954 &  3.526 & $-$     \\
$\beta_{0}$ &  2.250 & 0.119 & 18.899 & $<0.001$  \\
$\beta_{1}$ &  0.366 & 0.053 &  6.890 & $<0.001$  \\
$\beta_{2}$ &  0.586 & 0.051 & 11.490 & $<0.001$ \\
$\beta_{3}$ &  0.474 & 0.052 &  9.102 & $<0.001$ \\
$\beta_{4}$ & -0.322 & 0.063 & -5.112 & $<0.001$   \\
\hline
\end{tabular}
}
\end{table}
The randomized quantile residuals in Figure \ref{Apli2-resi}(a) shows the absence of
extra variability and atypical road, and Figure \ref{Apli2-resi}(b) shows a normal probability plot with simulated envelope for the randomized quantile residual.

\begin{figure}[h!]
\begin{minipage}[b]{0.48\linewidth}
\includegraphics[width=\linewidth]{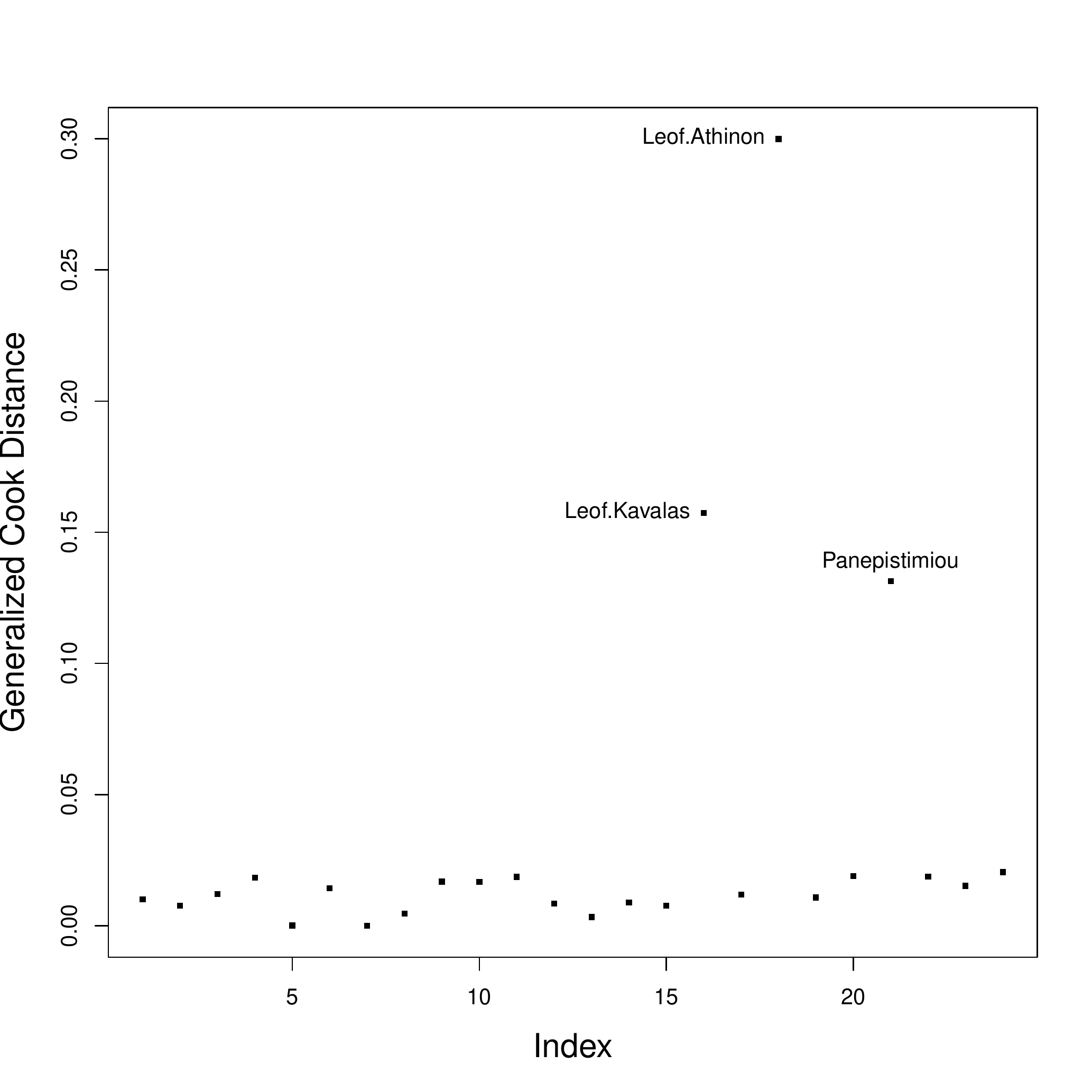}
\begin{center}
(a)
\end{center}
\end{minipage}
\hfill
\begin{minipage}[b]{0.48\linewidth}
\includegraphics[width=\linewidth]{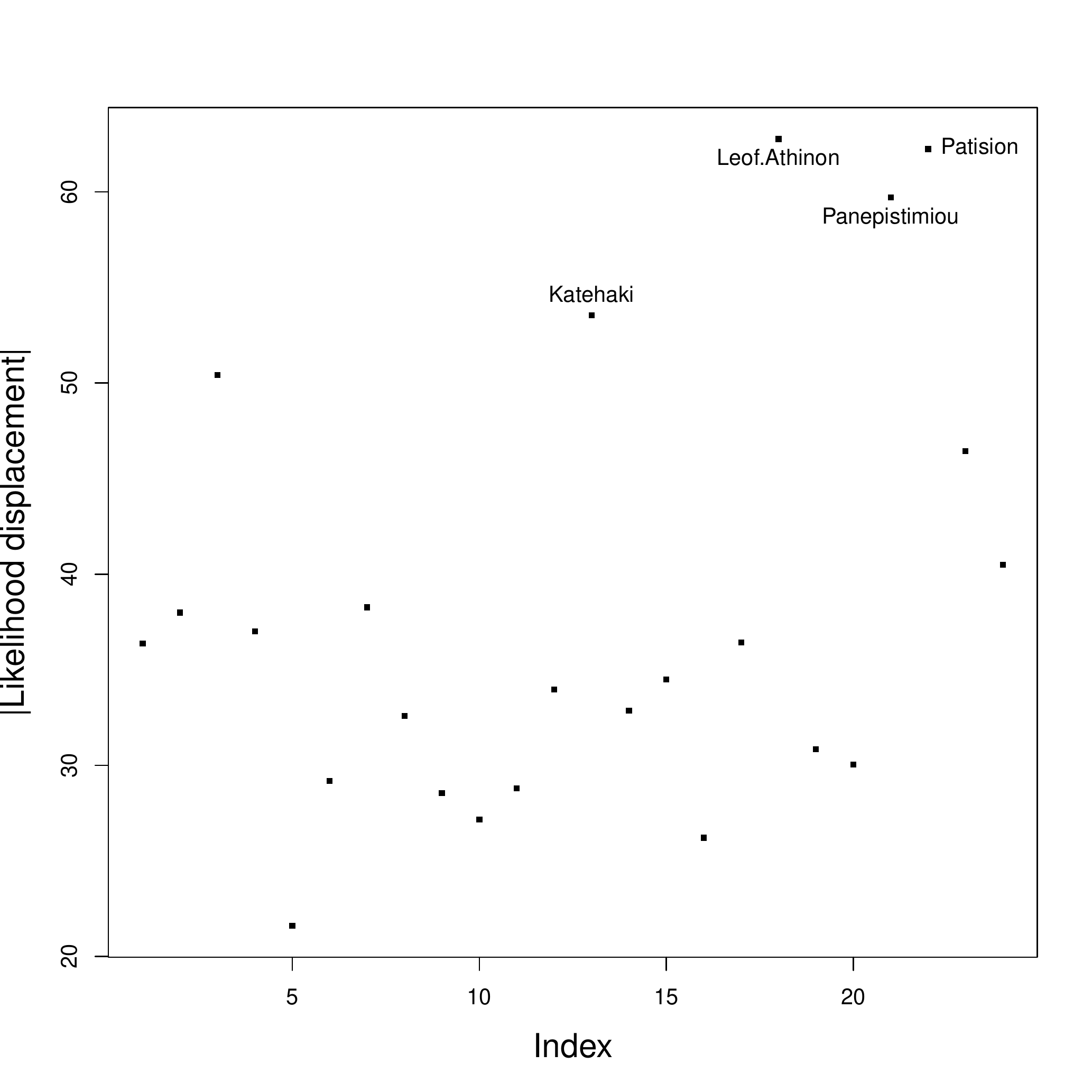}
\begin{center}
(b)
\end{center}
\end{minipage}
\caption{Index plot of (a) the generalized Cook distance and (b) the likelihood displacement for the number of  car accidents on central Athenian roads in 1989-1991.  }
\label{apli2-global}
\end{figure}

\begin{figure}[h!]
\begin{minipage}[b]{0.48\linewidth}
\includegraphics[width=\linewidth]{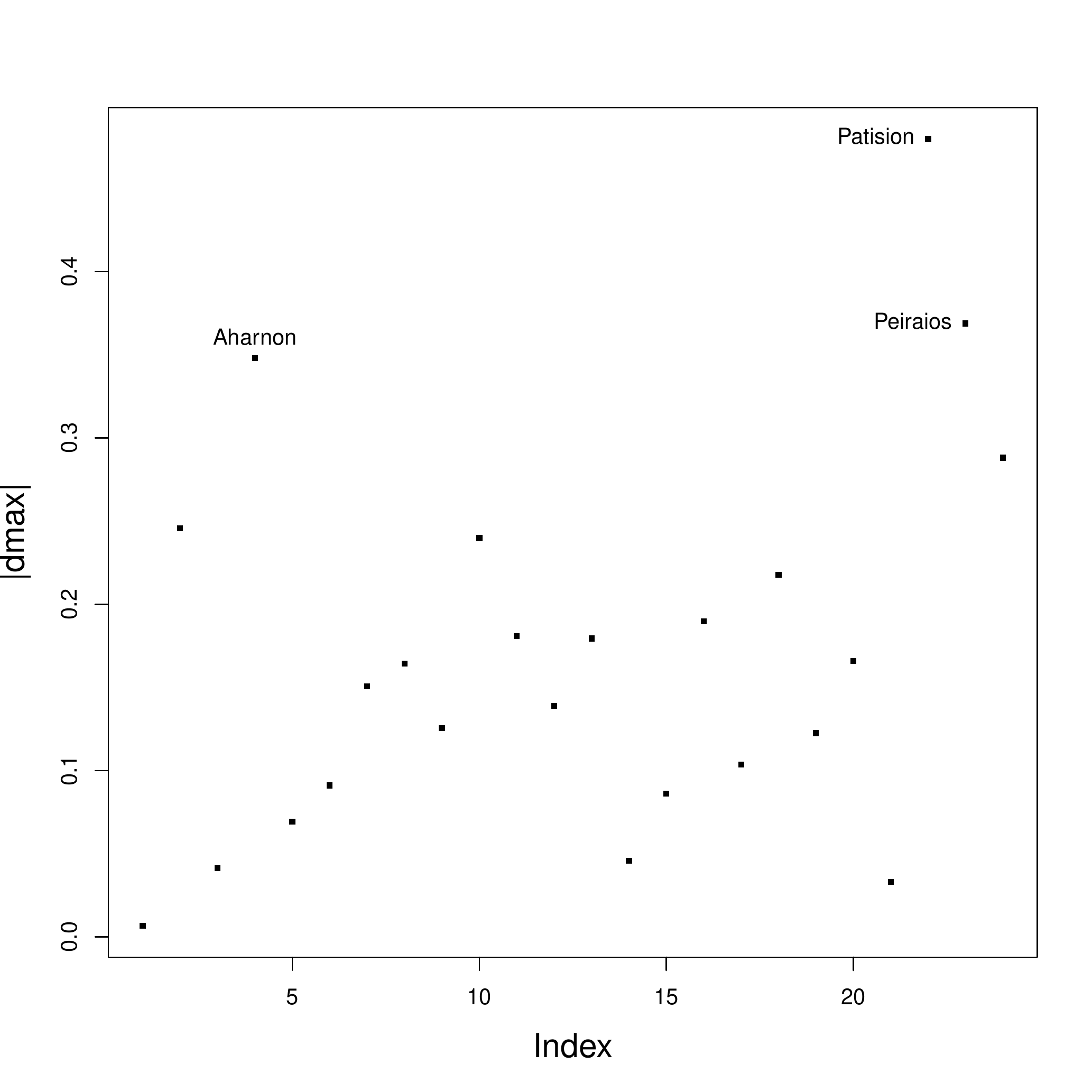}
\begin{center}
(a)
\end{center}
\end{minipage}
\hfill
\begin{minipage}[b]{0.48\linewidth}
\includegraphics[width=\linewidth]{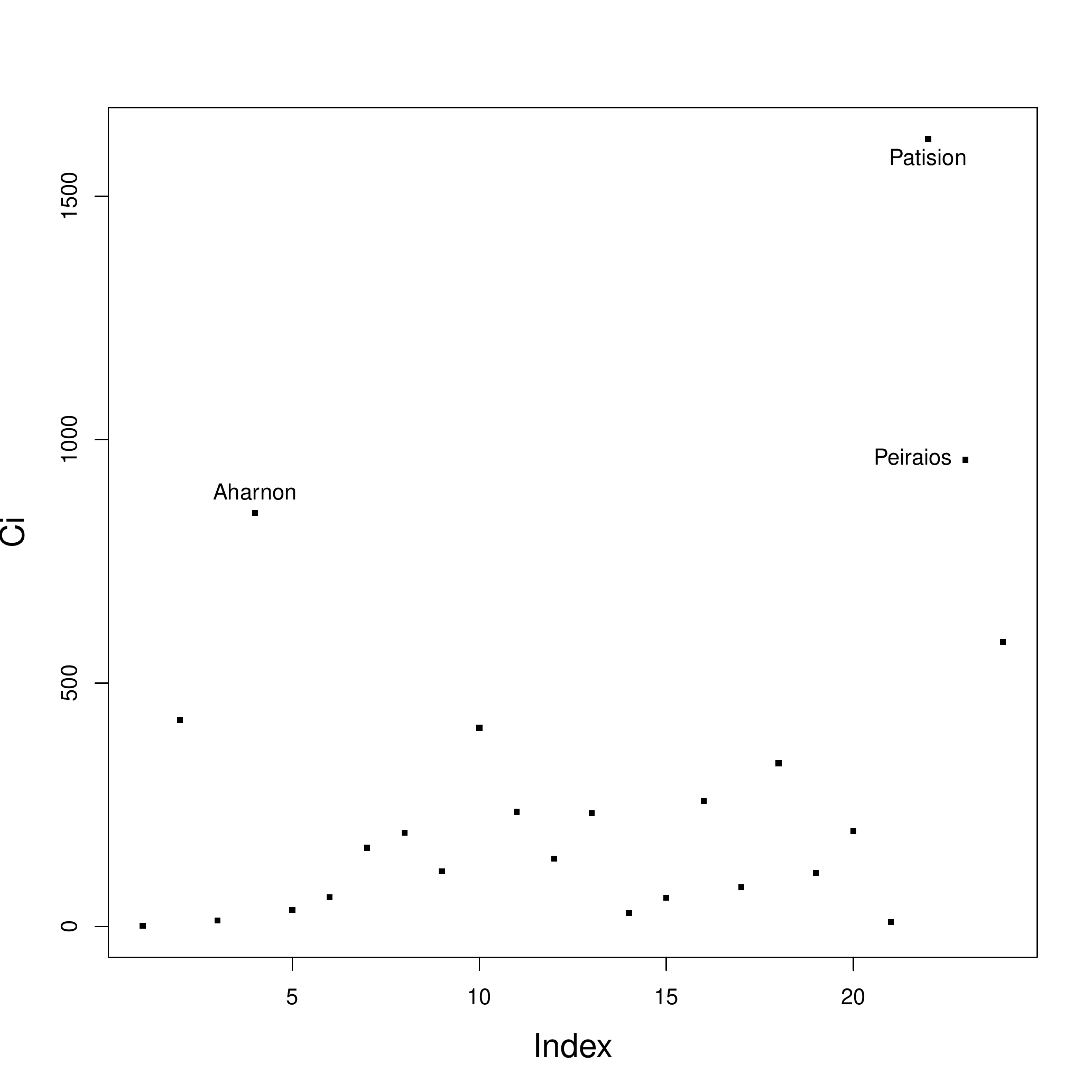}
\begin{center}
(b)
\end{center}
\end{minipage}
\caption{Index plot of $|\bm{d}_{max}|$ for the case weight perturbation scheme: (a) local influence and (b) total local influence for the number of  car accidents on central Athenian roads in 1989-1991.}
\label{local_apli2}
\end{figure}

The global influential graphic in Figure \ref{apli2-global}(a) indicates that $Leof.~Kavalas$, $Leof.~Athinon$ and $Panepistimiou$ can impact the ML estimates when they are removed. $Leof.~Kavalas$ (4,6,12,9,3) and $Panepistimiou$ (24, 58, 40, 36, 05)  are small roads, 2.0 and 1.1 kilometers long, respectively. $Leof.~Athinon$ (15,11,16,21,28) is a large road, 6.1 kilometers long. Furthermore, two roads that can impact the ML estimates are showed in Figure \ref{apli2-global}(b), $Patision$ (80,108,114,113,86) and $Katehaki$(2,3,27,24,7), which are 4.1 and 1.4 kilometers long, respectively. 
In Figure \ref{local_apli2}(a) and (b) are showed influential roads, $Peraios$ (86  89 109  90  49) and $Aharnon$ (44, 79, 91, 88, 33), which are 8.0 and 5.5 kilometers long, respectively. 

\begin{figure}[h!]
\begin{minipage}[b]{0.48\linewidth}
\includegraphics[width=\linewidth]{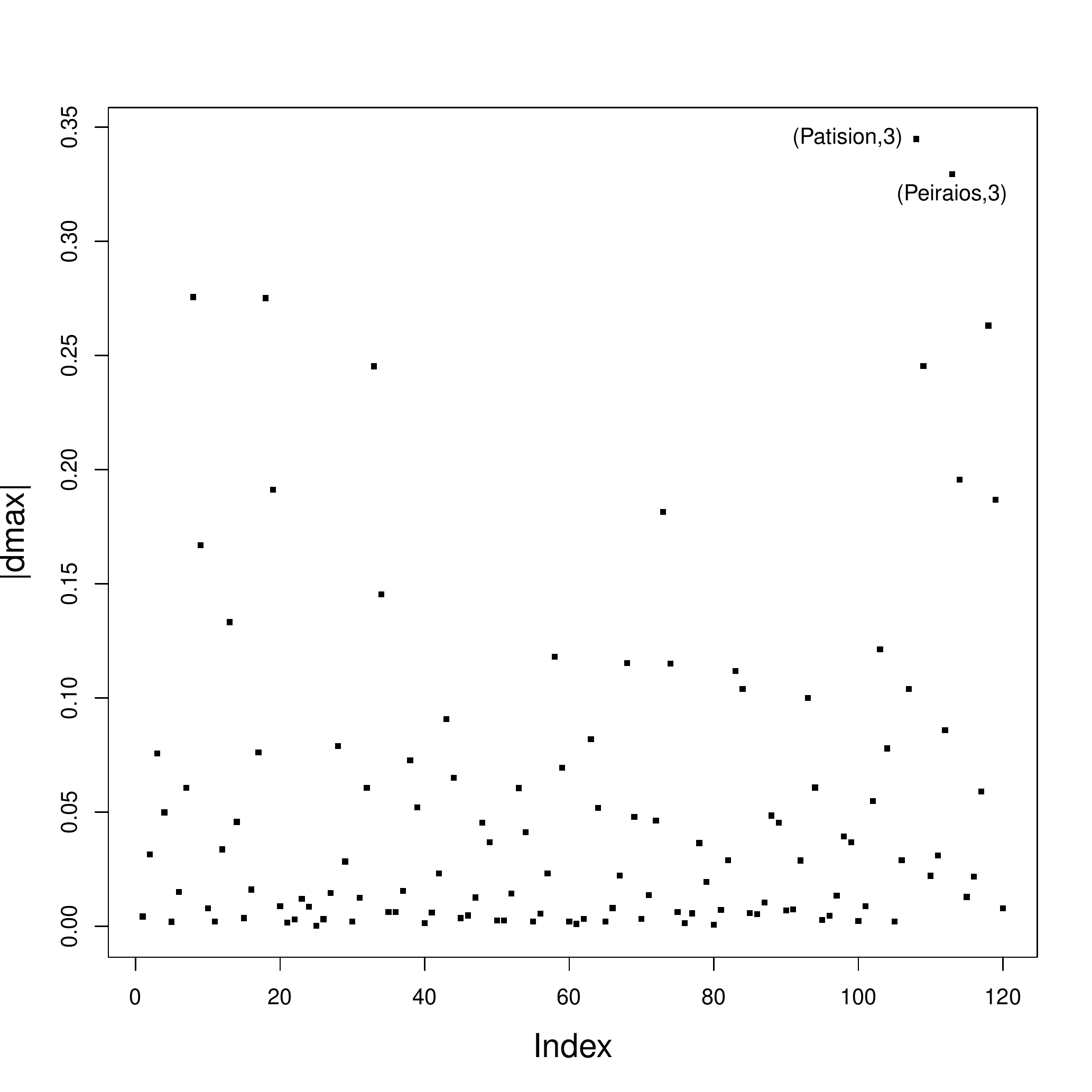}
\begin{center}
(a)
\end{center} 
\end{minipage}
\hfill
\begin{minipage}[b]{0.48\linewidth}
\includegraphics[width=\linewidth]{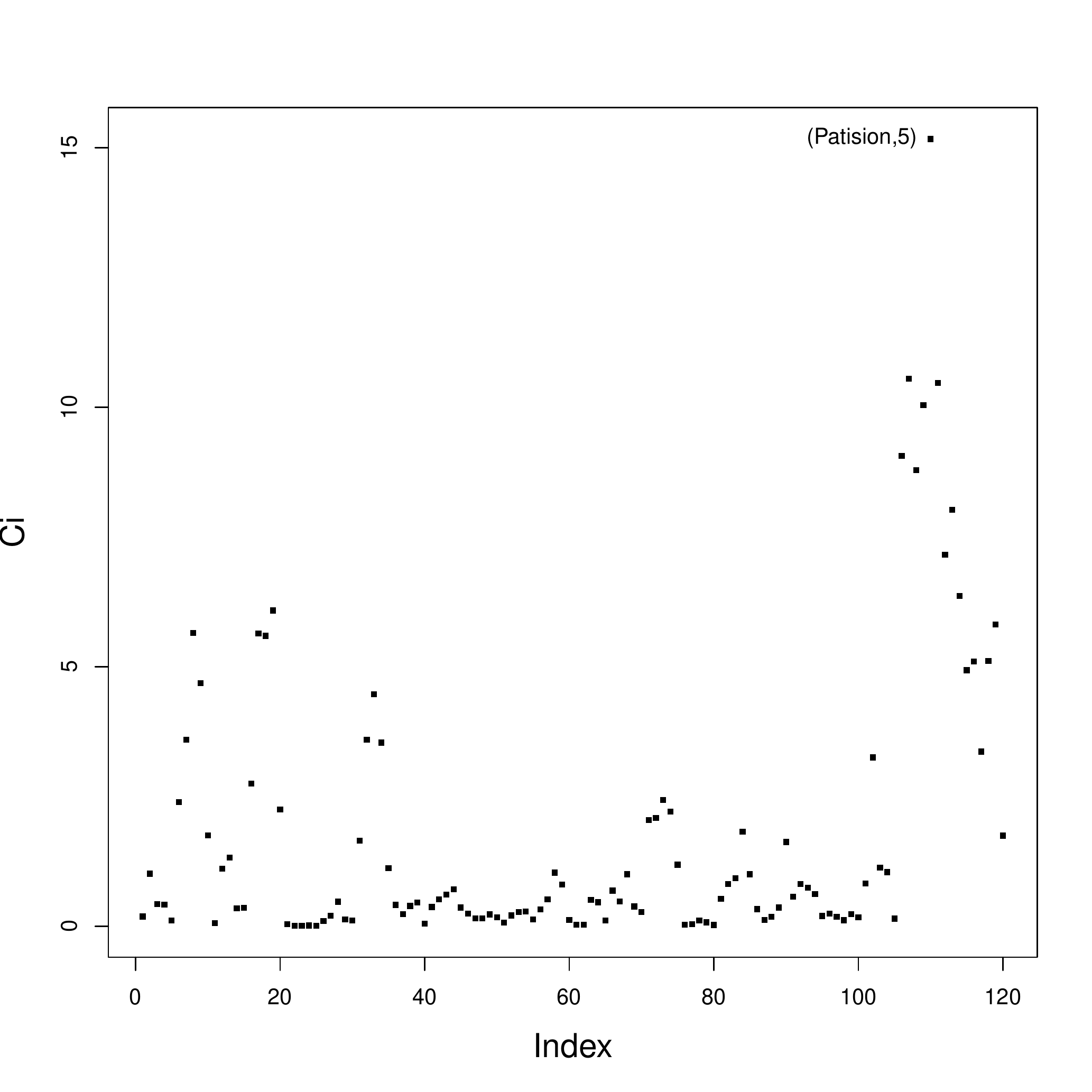}
\begin{center}
(b)
\end{center}
\end{minipage}
\caption{Index plot of $|\bm{d}_{max}|$ for the case weight perturbation scheme: (a) local influence and (b) total local influence on $j$th measurement of the number of car accidents for each atypical central Athenian road.}
\label{localtotal_obs}
\end{figure}
\begin{figure}[h!]
\begin{minipage}[b]{0.48\linewidth}
\includegraphics[width=\linewidth]{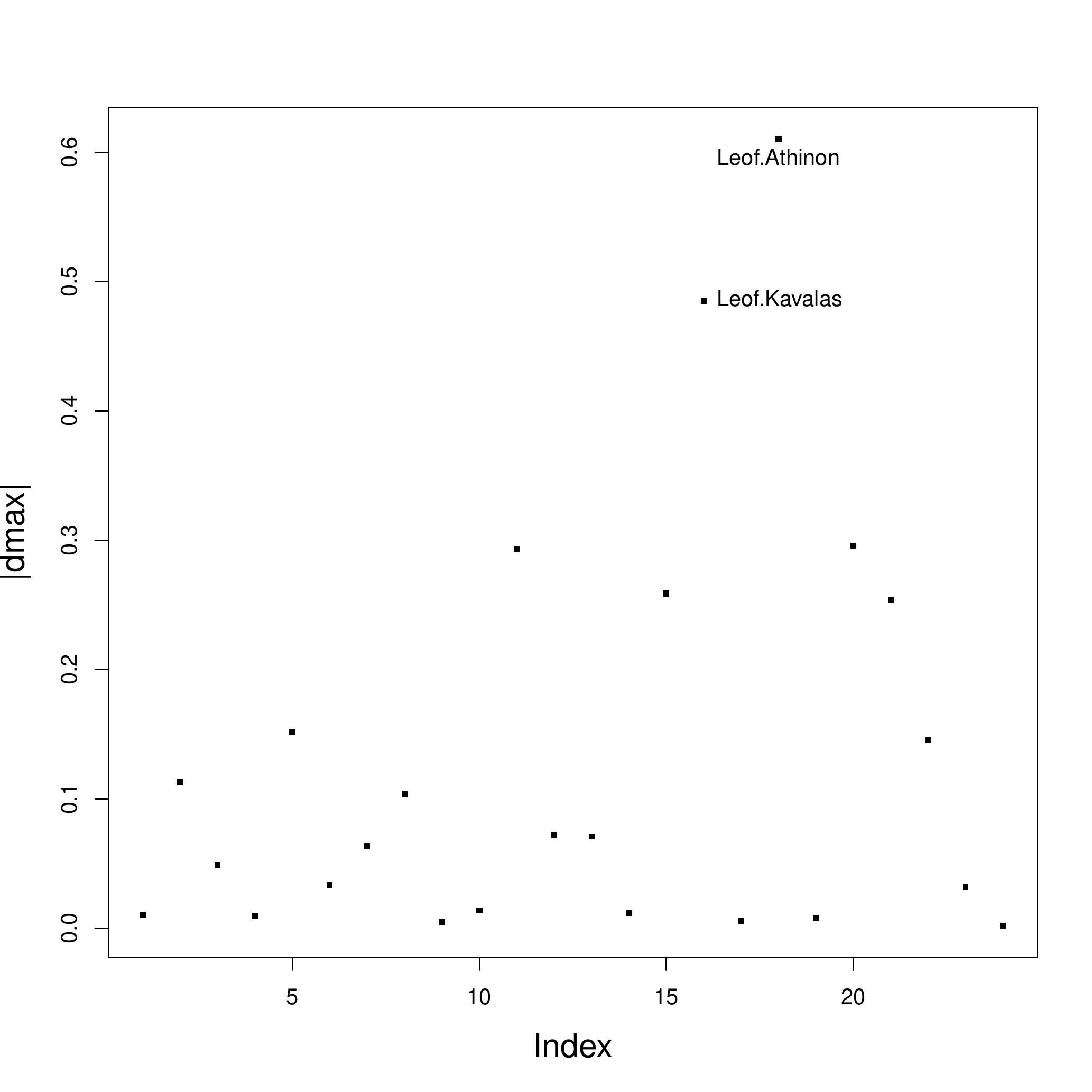}
\begin{center}
(a)
\end{center}
\end{minipage}
\hfill
\begin{minipage}[b]{0.48\linewidth}
\includegraphics[width=\linewidth]{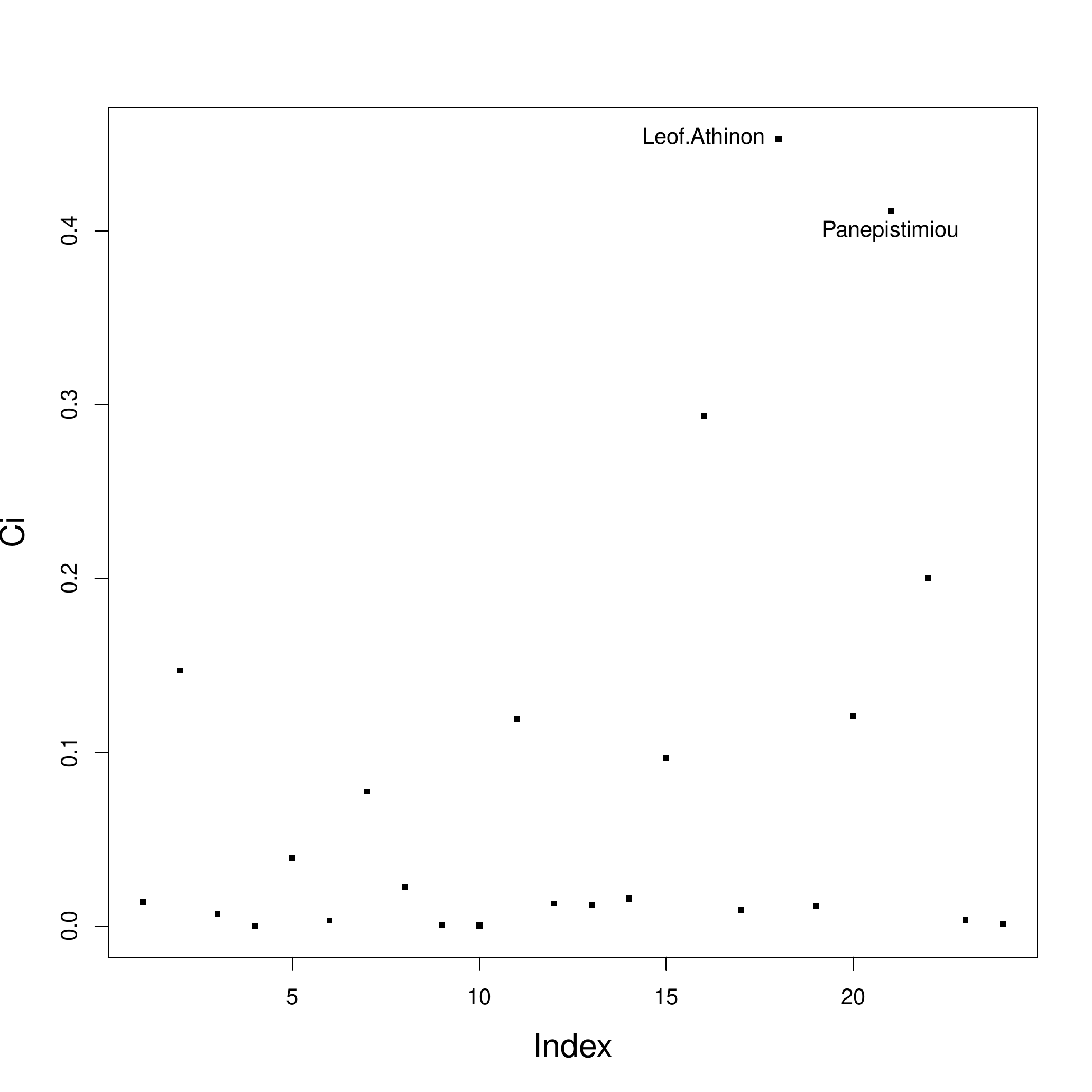}
\begin{center}
(b)
\end{center}
\end{minipage}
\caption{Index plot of $|\bm{d}_{max}|$ for the dispersion perturbation scheme, (a) local influence and (b) total local influence for the number of  car accident on central Athenian roads of 1989-1991.}
\label{local:dispesao}
\end{figure}
Finally, the percentage relative deviations ($PRD$) are calculated and we verify that the most significant changes in the ML estimates are associated with the removal of the influential roads $Patision$ and $Peiraios$. The results in Table \ref{deleteApli2} show that the estimates are affected when $Patision$ and $Peiraios$ are removed simultaneously. 
\begin{table}[h!]
\centering {
\caption{Parameter estimates, standard errors (Std. Error), z-values, $p$-values, and $PRD(\%)$ when $Patision$ road is deleted from the accidents data.}
\label{deleteApli2}
\begin{tabular}{c|c|rrrrr}
\hline
Dropping & Parameter & Estimate & Std. Error & z-value & $p$-value  &  $PRD(\%)$ \\
\hline
$Patision$ & $\phi$      &  3.507 & 1.021 & $-$    &  $-$   &  $-$ \\
& $\beta_{0}$ &  2.199 & 0.120 & 18.291 & $<0.001$  &  2.27\\
& $\beta_{1}$ &  0.376 & 0.057 &  6.596 & $<0.001$  &  -2.65\\
& $\beta_{2}$ &  0.617 & 0.054 & 11.331 & $<0.001$  &  -5.36\\
& $\beta_{3}$ &  0.492 & 0.055 &  8.831 & $<0.001$  &  -3.86\\
& $\beta_{4}$ & -0.399 & 0.069 & -5.765 & $<0.001$  &  -23.97\\
\hline
$Peiraios$ & $\phi$      &  3.235 & 0.934 & $-$    &  $-$   &  $-$ \\
& $\beta_{0}$ &  2.215 & 0.124 & 17.795 & $<0.001$  &  1.54\\
& $\beta_{1}$ &  0.413 & 0.057 &  7.249 & $<0.001$  &  -12.63\\
& $\beta_{2}$ &  0.634 & 0.056 & 11.605 & $<0.001$  &  -8.23\\
& $\beta_{3}$ &  0.531 & 0.055 &  9.534 & $<0.001$  &  -11.98\\
& $\beta_{4}$ & -0.287 & 0.067 & -4.257 & $<0.001$  &   10.91\\
\hline
$Peiraios~and~Patision$ & $\phi$      &  3.366 & 0.934 & $-$    &  $-$   &  $-$ \\
												& $\beta_{0}$ &  2.154 & 0.126 & 17.070 & $<0.001$  &  4.24\\
												& $\beta_{1}$ &  0.432 & 0.062 &  7.001 & $<0.001$  &  -17.95\\
												& $\beta_{2}$ &  0.678 & 0.059 & 11.491 & $<0.001$  &  -15.76\\
												& $\beta_{3}$ &  0.561 & 0.060 &  9.319 & $<0.001$  &  -18.47\\
												& $\beta_{4}$ & -0.370 & 0.075 & -4.924 & $<0.001$  &   -14.91\\
\hline
\end{tabular}
}
\end{table}
\section{Discussion}
\label{conclu}
In this study, we developed diagnostic tools for the MNBR model derived from the Poisson mixed model where the GLG distribution is assumed for the random effect. Simulation results exhibited the features of the MNBR model and its association with the hierarchical model, which was essential for explaining that the asymptotic consistency of the estimators of the regression coefficients and the dispersion parameter depends on the asymmetry of the GLG distribution. As expected, the MNBR model provides inconsistent estimates of the asymptotic variance of the ML estimators when the covariance matrix of the response variable is misspecified. 
The randomized quantile residuals can be used to assess possible departures of the data from the MNBR model assumptions. 
Following the approaches of \citet{Cook77,Cook} and \citet{Lesafre}, global and local measures for the MNBR model were derived and implemented in the authors’ \texttt{MNB} package. The application of the \texttt{MNB} package to two data sets was presented and it was shown that  the asymmetric behavior of the empirical distributions of the individual profiles can indicate the need to use the multivariate model to better fit these profiles. The proposed methodology was helpful for identifying outlying subjects that matter in the MNBR model over time and handle the overdispersion phenomenon.  The code for installing the \texttt{MNB} package is presented in the Appendix. 
\section*{Acknowledgements}
The work of the fourth author is partially funded by CNPq, Brazil. We also thank anonymous referees for constructive comments and suggestions.

\setlength{\bibsep}{0.0pt}
\bibliographystyle{natbib} 


\section*{Appendix}
\section*{1. The score vector and the observed Fisher information matrix}

The score vector $\bm{U}_{\theta}=(\bm{U}_{\beta}^{\top}, U_{\phi})^{\top}$ is obtained by deriving the log-likelihood function (2) with respect to the parameters $\bm{\beta}$ and $\phi$, respectively. Thus, 
\begin{align}
\bm U_{\beta} & =\sum_{i=1}^{n}\bm{X}_i^{\top}(\bm{y}_i-a_i\bm{\mu}_i) \quad \text{and} \nonumber\\
U_{\phi} & =  \sum_{i=1}^{n} \bigg\{ \psi(\phi + y_{i+})-
                   \psi(\phi)+\log \left(\frac{\phi}{\phi + \mu_{i+}}\right) +
\left(\frac{\mu_{i+} - y_{i+}}{\phi + \mu_{i+}}\right) \bigg\}\label{ape:scphi},
\end{align}
where $a_i=(\phi + y_{i+})/(\phi+\mu_{i+}),$ $\psi(\cdot)$ is the digamma function, and $\bm{X}_{i}$ is an $m_{i} \times p$ matrix with row $\bm{x}^{\top}_{ij}$ for $i=1,\ldots,n$ and $j=1,\ldots,m_i.$  However, using the fact that 
 $\Gamma(\phi + y_{i+} )/\Gamma(\phi) = \phi(\phi +1)(\phi + 2)\ldots (\phi + y_{i+}-1)$ in 
(\ref{ape:scphi}), $U_{\phi}$ takes the following form
$$
U_{\phi}=\sum_{i=1}^{n}\left\{\sum_{j=0}^{y_{i+ - 1}}(j + \phi)^{-1} - \frac{y_{i+}}{\phi + \mu_{i+}} - \log(1 + \phi^{-1}\mu_{i+}) + \frac{\mu_{i+}}{(\phi + \mu_{i+})}\right\}.
$$
The observed information matrix is obtained by deriving the score vector with respect to $\bm{\beta}$ and $\phi$. Thus, 
$$-\ddot{\ell}(\bm{\theta}) =
\begin{bmatrix}
-\ddot{\ell}_{\beta\beta} & -\ddot{\ell}_{\beta\phi}\\ -\ddot{\ell}_{\phi\beta} & -\ddot{\ell}_{\phi\phi},
\end{bmatrix}
$$
where
$$
\ddot{\ell}_{\beta\beta}=-\sum_{i=1}^{n}\frac{(\phi + y_{i+})}{(\phi + \mu_{i+})}\sum_{j=1}^{m_i}\bm{x}_{ij}\bm{x}^{\top}_{ij}\mu_{ij} + \sum_{i=1}^{n}\frac{(\phi + y_{i+})}{(\phi + \mu_{i+})^2}\sum_{j=1}^{m_i}\bm{x}_{ij}\mu_{ij}\sum_{j=1}^{m_i}\bm{x}^{\top}_{ij}\mu_{ij},
$$

$$
\ddot{\ell}_{\beta\phi}=\ddot{\ell}_{\phi\beta}=-\sum_{j=1}^{m_i}x_{ijk}\mu_{ij}\frac{(\mu_{i+} - y_{i+})}{(\phi+\mu_{i+})^2} \nonumber \ \ {\rm and}
$$

$$
\ddot{ \ell}_{\phi\phi}=-\sum_{i=1}^{n}\left\{\sum_{s=0}^{y_i^{*}}\frac{1}{(s+\phi)^2} + \frac{\phi^{-1}\mu_{i+}}{(\phi + \mu_{i+})} - \frac{\mu_{i+}}{(\phi + \mu_{i+})^2} - \frac{y_{i+}}{(\phi + \mu_{i+})^2}\right\}.
$$

\section*{2. How to install and use the \texttt{MNB} package}

We present the \texttt{R} platform code used to install the \texttt{MNB} package.\\

\begin{lstlisting}
require(devtools)
devtools::install_github("carrascojalmar/MNB")
\end{lstlisting}

\subsection*{2.1. Codes for Seizures data}

\begin{lstlisting}
require(MNB)
data(seizures)

#
star <-list(phi=1, beta0=1, beta1=1, beta2=1, beta3=1)
mod <- fit.MNB(formula=Y ~ trt + period +
trt:period + offset(log(weeks)), star=star, dataSet=seizures,tab=FALSE)
mod

#
par <- mod$par
names(par)<-c()
res.q <- qMNB(par=par,formula=Y ~ trt + period + trt:period +
offset(log(weeks)),dataSet=seizures)

plot(res.q,ylim=c(-3,4.5),ylab="Randomized quantile residual",
xlab="Index",pch=15,cex.lab = 1.5, cex = 0.6, bg = 5)
abline(h=c(-2,0,2),lty=3)

#
envelope.MNB(formula=Y ~ trt + period + trt:period +
offset(weeks),star=star,nsim=21,n.r=6,
dataSet=seizures,plot=TRUE)

#
global.MNB(formula=Y ~ trt + period +
trt:period + offset(log(weeks)),star=star,dataSet=seizures,plot=TRUE)

#
local.MNB(formula=Y ~ trt + period + trt:period + offset(log(weeks)),star=star,dataSet=seizures,
schemes="weight",plot=TRUE)

\end{lstlisting}

\subsection*{2.2. Codes for Accident data}

\begin{lstlisting}

require(MNB)
accident <- read.table("...\\accident.txt",h=TRUE)
temp <- apply(accident[,2:6],1,t)
Y <- matrix(temp,ncol=1)
length.x <- rep(accident$length,each=5)
ind <- rep(1:24,each=5)
year <- as.factor(rep(1:5,24))
accident2 <- data.frame(Y,year,length.x,ind)

#
star=list(phi=10, beta1=2, beta2=0.4, beta3= 0.6, beta4=0.5, beta5=-0.3)
mod <- fit.MNB(Y ~ year + offset(log(length.x)),star=star,dataSet=accident2,tab=FALSE)
mod

#
par <- mod$par
names(par)<-c()
qMNB(par=par,formula=Y ~ year + offset(log(length.x)),dataSet=accident2)

#
envelope.MNB(formula=Y ~ year + offset(log(length.x)),star=star,nsim=21,n.r=6,
dataSet=accident2,plot=TRUE)

#
global.MNB(Y ~ year + offset(log(length.x)),star=star,dataSet=accident2,plot=TRUE)

#
local.MNB(Y ~ year + offset(log(length.x)),star=star,dataSet=accident2,
schemes="weight",plot=TRUE)

\end{lstlisting}

\end{document}